\documentclass[reprint,aps,prl,superscriptaddress,english,nolongbibliography]{revtex4-1}

\usepackage{graphicx,dcolumn,braket}
\usepackage[table,xcdraw,dvipsnames]{xcolor}
\usepackage[colorlinks=true,citecolor=blue,urlcolor=blue,linkcolor=blue]{hyperref}
\usepackage[caption=false,subrefformat=parens,labelformat=empty]{subfig}
\usepackage{amsmath,amssymb,bm,mathrsfs,amsfonts}
\usepackage{verbatim}
\usepackage{physics}

\begin{document}
\title{Renormalization group for Anderson localization on high-dimensional lattices}

\author{Boris L.\ Altshuler}
\affiliation{Physics Department, Columbia University, 538 West 120th Street, New York, New York 10027, USA}
\author{Vladimir E.\ Kravtsov}
\affiliation{ICTP, Strada Costiera 11, 34151, Trieste, Italy}

\author{Antonello Scardicchio}
\affiliation{ICTP, Strada Costiera 11, 34151, Trieste, Italy}
\affiliation{INFN Sezione di Trieste, Via Valerio 2, 34127 Trieste, Italy}

\author{Piotr Sierant}
\affiliation{ICFO-Institut de Ci\`encies Fot\`oniques, The Barcelona Institute of Science and Technology, Av. Carl Friedrich Gauss 3, 08860 Castelldefels (Barcelona), Spain}

\author{Carlo Vanoni}
\email{cvanoni@sissa.it}
\affiliation{SISSA -- International School for Advanced Studies, via Bonomea 265, 34136, Trieste, Italy}
\affiliation{INFN Sezione di Trieste, Via Valerio 2, 34127 Trieste, Italy}

\begin{abstract}
    We discuss the dependence of the critical properties of the Anderson model on the dimension $d$ in the language of $\beta$-function and renormalization group recently introduced in Ref.~\cite{vanoni2023renormalization} in the context of Anderson transition on random regular graphs. We show how in the delocalized region, including the transition point, the one-parameter scaling part of the $\beta$-function for the fractal dimension $D_{1}$ evolves smoothly from its $d=2$ form, in which $\beta_2\leq 0$, to its $\beta_\infty\geq 0$ form, which is represented by the random regular graph (RRG)  result. We show how the $\epsilon=d-2$ expansion and the $1/d$ expansion around the RRG result can be reconciled and how the initial part of a renormalization group trajectory governed by the irrelevant exponent $y$ depends on dimensionality. We also show how the irrelevant exponent emerges out of the high-gradient terms of expansion in the nonlinear sigma-model and put forward a conjecture about a lower bound for the fractal dimension. The framework introduced here may serve as a basis for investigations of disordered many-body systems and of more general non-equilibrium quantum systems.
\end{abstract}

\maketitle

Despite extensive, centuries-long research in the field of statistical mechanics, the mechanisms underlying the process of thermalization are still not fully understood. The analog of the ergodic hypothesis of classical mechanics in quantum mechanical systems, and its validity in presence of quenched disorder, present several counter-intuitive aspects which attracted the interest of the scientific community working on foundations and applications of statistical mechanics. It is known that when the interaction between the particles can be neglected and the disorder is sufficiently strong, the system undergoes a transition from an ergodic to a localized, Anderson insulator phase \cite{Anderson1958absence,ohtsuki1999review,evers2008anderson} which has no counterpart in classical mechanics. The properties of the corresponding Anderson transition are understood to depend upon the physical dimension of space $d$ and, as the $d$ increases indefinitely, those properties have been a subject of growing interest in the recent past \cite{Tarquini2017critical,baroni2023corrections, tikhonov2016fractality,tikhonov2016Anderson,bera2018return,sierant2023universality,Lemariè2022critical}.

In part, this is due to the interest in the complementary case, in which the elementary excitations of the system {\it cannot} be thought of as non-interacting particles, and interaction needs to be considered in the analysis. The analog of Anderson localization, in this case, is the subject of Many-Body Localization (MBL) \cite{Basko06,nandkishore2015many,Abanin2019colloquium}, where the system develops local integrals of motion \cite{ros2015integrals,Imbrie17} and transport is suppressed~\cite{vznidarivc2016diffusive}. The connection between MBL and the problem of Anderson localization occurs when thinking of the latter on infinite-dimensional lattices, or expander graphs, such as trees, and regular random graphs (RRG) \cite{altshuler1997quasiparticle,de2014anderson, tikhonov2021AndersonMBL} which have much in common with the structure of the Hilbert space of interacting systems. Moreover, some of the difficulties in interpreting the numerical data supporting MBL (see for example \cite{vsuntajs2020quantum,panda2020can,abanin2021distinguishing,Sierant20Thouless, sierant2022challenges, sierant24MBLreview}) have very much in common with the difficulties of interpreting the numerical data of the Anderson model on the RRG (where there is no doubt about the existence of the transition \cite{abou1973selfconsistent}).

There is another reason for the current interest in the Anderson transition on expander graphs. The success of the one parameter scaling theory~\cite{abrahams1979scaling} supported by the supersymmetric (or replica) sigma model~\cite{Wegner1987,zirnbauer1986anderson} in understanding the Anderson problem at low dimensions $d=1$ and $d=2$ as well as in the description of the Anderson transition for $d=2+\epsilon$ created a long lasting impression that we qualitatively understand the properties of the transition at all $d>2$ , in particular at $d=3$. However, the absence of an obvious upper critical dimension and the failure of $\epsilon$ expansion around $d=2$ dimensions to fit the numerically found exponents \cite{ueoka2014dimensional} at $\epsilon=1$ ($d=3$), despite going to five loops in the sigma model \cite{hikami1992localization}, remains puzzling. 

In this paper that we build upon the work that some of us did in Ref.~\cite{vanoni2023renormalization}, we show how a scaling theory (a modern form of the one presented in Ref.~\cite{abrahams1979scaling}) can explain the numerics on the statistical properties of wave functions and spectrum. We also show, connecting to our work Ref.~\cite{vanoni2023renormalization}, that the irrelevant corrections, in the RG sense, to the one-parameter scaling evolve in the limit of infinite dimensions to give rise to a topologically different RG flow.

To set the stage we briefly recall the original version of the scaling theory of Ref.~\cite{abrahams1979scaling}. We then reformulate it in terms of an observable which has clear meaning for a much broader class of the models than the dimensionless Thouless conductance and at the same time is easier to access numerically. This observable is the \emph{finite-size} fractal dimension $D$ of the eigenstates.
While its behavior can be predicted analytically in some regimes (as in the deep ergodic and localized regimes), we have to rely on numerical results for the properties near the critical point. We show that our framework is compatible with the existing numerical observations and clearly describes the model at any $d$. 
This is achieved by matching the known exact results in $2d$ -- and perturbations away from it in the $\epsilon$-expansion framework -- to the results on RRGs, that, as we argue, coincides with the ones in the limit $d \to \infty$.

\section*{The Anderson Model and a scaling theory of the conductance}

The Anderson model originally introduced in Ref.~\cite{Anderson1958absence} describes a single quantum particle (whose statistics is thus not important) hopping on a given graph $\Lambda$ in the presence of onsite random fields. In the case of a $d$-dimensional cubic lattice of linear size $L$, that we study here, the volume of the system (i.e. the number of sites) is $N=L^d$. The Hamiltonian operator defining the model is
\begin{equation}
    H=- J \sum_{\langle i, j \rangle \in\Lambda}\dyad{i}{j}+\mathrm{h.c.}+\sum_{i\in \Lambda}\epsilon_i\dyad{i}{i}.
\end{equation}
In the above expression, $\langle \cdot \rangle$ represent nearest neighbor sites on the graph $\Lambda$ and the on-site energies $\epsilon_i$ are distributed uniformly according to the box distribution $g(\epsilon) = \theta(|\epsilon| - W/2)/W$. We choose the hopping rate as the unit of energy, $J=1$. The eigenstates $\psi_n$ have the energy $E_{n}$ found from the Schr\"odinger equation $H \ket{\psi_n}=E_n \ket{\psi_n}$.

It is known that the model can have a transition from diffusive/ergodic to localized  phase as the strength  of disorder $W$ increases. The location of such transition (i.e. the critical value of $W = W_c$) strongly depends on the structure of the lattice $\Lambda$, while the critical exponents at the transition are universal and depend only on the lattice dimensionality $d$.   

The seminal work~\cite{abrahams1979scaling} discuses  the dependence of dimensionless conductance on the system size and the strength of disorder $W$. The main result of the paper which determined the development of the field for decades, was a formulation of the single-parameter scaling. It stated that   the evolution of conductance with the system size $L$ depends only on the conductance itself and not on $L$ and $W$ separately. This allowed to uncover the crucial role of the lattice dimensionality $d$ and predict the absence of delocalized states for $d=1$ {\it in the thermodynamic limit}, as well as the existence of the localization/delocalization transition for $d>2$.  
   
In Ref.~\cite{abrahams1979scaling} the main observable is the dimensionless conductance  $g(L)$, where $L$ is the linear size of the system. $g(L)$ is defined as the ratio: 
\begin{equation*}
    g(L) =\frac{2 \pi \hbar}{e^2}\sigma L^{d-2}
\end{equation*}
where   $\sigma$ is the conductivity.  The mathematical formulation of the single parameter scaling is given by:
\begin{equation}
    \frac{d \ln g(L)}{d \ln L} = \beta(g(L)),
\end{equation}
where $\beta(g)$ is the parameter-free $\beta$-function.  

Already from the definition of $g(L)$, it is easy to see that in the developed metallic regime (where $\sigma$ is $L$-independent), the $\beta$-function is a positive constant $\beta(g)=(d-2)$. In the deep insulator regime $\sigma\sim {\rm exp}(-L/\xi)$, the $\beta$-function is $(-L/\xi)=\ln(g)$ is negative. A continuous interpolation between these two regimes for $d>2$ inevitably leads to the unstable fixed point $g_{c}$ such that $\beta(g_{c})=0$ which corresponds to the localization/delocalization transition. If for small system sizes the initial value is $g_{0}>g_{c}$, the conductance $g(L)$ increases with $L$ driving the system to the metallic regime, while at $g_{0}<g_{c}$ the conductance decreases with $L$ and, eventually, the system reaches the deep insulating regime. In contrast to this scenario, if $d<2$ (e.g. $d=1$) the $\beta$-function is everywhere negative and the metallic behavior is not possible. The case of the two-dimensional lattice is special, as at $g\rightarrow\infty$ we have $\beta(g)\rightarrow 0$ (e.g. $d=2$ is a critical dimensionality). A more careful perturbative study in $1/g$ shows that for disordered potentials without spin-orbit interaction this limit is reached from below, so that the simplest assumption of a monotonic $\beta$-function leads to the conclusion that $\beta(g)<0$ everywhere, e.g. on the absence of delocalized states for $d=2$. Expanding the $\beta$-function around $g=g_{c}$ it is possible to determine some critical properties, such as the exponent $\nu = 1/s$, where $s$ is the logarithmic slope of the $\beta$-function at the critical point $\beta(g) = s \ln(g/g_c)$. For more details, we refer to~\cite{abrahams1979scaling}.

\section*{$\beta$-function for ‘modern' observables}
\subsection*{Numerically accessible scaling variables}

The Anderson localization transition affects most observable properties of the system. The onset of the localized phase can be spotted not only from the absence of transport (as in the original work by Anderson~\cite{Anderson1958absence}), but also through properties of the spectrum and statistics of eigenfunctions.  
Modern libraries for high-performance computing make spectral statistics and eigenfunctions statistics  readily accessible alternative of dimensionless conductance.

A basic statistics of the eigenfunctions  is the Shannon entropy:
\begin{equation}
\label{eq:vN_entr}
    S_n =-\sum_{i}\psi^2_n(i)\ln(\psi^2_n(i)).
\end{equation}

The derivative $D(L)$ with respect to $\ln L$, or the logarithm of volume $\ln N=\ln(L^{d})$, of the average value $S=\langle S_{n}\rangle$ of the  Shannon entropy is a fundamental quantity. In the limit $N\rightarrow\infty$ it gives the fractal dimension of the eigenfunction support set, which is equal to:
\begin{equation}\label{D-de}
D\equiv\lim_{N\rightarrow \infty}\frac{d S}{d\ln N}=
\left\{
\begin{matrix} 
    1, &  \mathrm{ergodic~states} \cr
    <1, & \mathrm{(multi)fractal~states} \cr 
    0,& \mathrm{localized~states} 
\end{matrix}
\right.
\end{equation} 
 
The special role of $D$ is seen from its connection with the spectral property of level compressibility defined as $\chi = \langle \delta n^2 \rangle/\langle n \rangle$, where $n$ is the number of energy levels in a given energy window and the average is over different positions of the energy window and over disorder realizations. It was shown~\cite{Chalker1996Spectral,Bogomolny2011Integrable,Bogomolny2011Eigenfunction}  that $\chi \approx (1-D)$ \footnote{for weak multifractality $\chi\ll 1$ one has   $1-D\approx (d-d_{q})/(q d)$ for all $q\geq 1$, so that $\chi\approx (d-d_{2})/2d$, see  \cite{Chalker1996Spectral,evers2008anderson} details and definitions.}.  

In view of a fundamental role of $D$ we choose, instead of the conductance $g(L)$, the scaling variable $D(L)$ defined in analogy with Eq.(\ref{D-de}) but {\it for a finite $L$}:
\begin{equation} \label{D_L_def}
D(L)=\frac{d S(L)}{d\ln N},
\end{equation}
where $S(L)$ is the average Shannon entropy in a system of size $L$. 

From the above discussion, it is clear that $D(L)$ is intimately related to spectral statistics, such as
 $r$-parameter introduced in Ref.~\cite{oganesyan2007localization} and defined as follows:
\begin{equation}
\label{eq:r-par}
    r=\frac{1}{N-2}\left\langle \sum_{n=1}^{N-2}\frac{\min(\Delta E_n,\Delta E_{n+1})}{\max(\Delta E_n,\Delta E_{n+1})}\right\rangle,
\end{equation}
where $\Delta E_{n}=E_{n+1}-E_{n}$ is the gap between the neighboring eigenvalues.
When $E_n$'s are eigenvalues of a real Hamiltonian, the average of $r$ takes values between $r_{\mathrm{GOE}} \simeq 0.5307$ and $r_{\mathrm{P}} = 2 \ln 2 -1 \simeq 0.386$. When $r = r_{\mathrm{GOE}}$ the spectrum behaves according to the predictions of random matrix theory (Gaussian orthogonal ensemble) and we expect the system dynamics to be ergodic. If instead $r = r_{\mathrm{P}}$, the energy levels are distributed independently (absence of level repulsion) and ergodicity is broken. Across the Anderson transition, the value of $r$ goes from $r_{\mathrm{GOE}}$ at small $W$ to $r_{\mathrm{P}}$ at large $W$. 

We would like to mention here an important difference between the $r$-statistics and the spectral compressibility $\chi$ that is directly related to $D(L)$. The point is that the former is defined at a small energy scale of the order of the mean level spacing $\delta$, while the latter (and presumably also $D(L)$) knows about level correlations at a scale much larger than $\delta$. This is important for sensing the multifractal-to-ergodic transition which in some cases does not show up in the $r$-statistics, as it happens, e.g. in the Rosenzweig-Porter random matrix model \cite{RP15}.

For this reason, we choose in this paper $D(L)$ as the scaling parameter
 in the RG equation:
\begin{equation}\label{RG_general}
\frac{d \ln D(L)}{d\ln N}=\beta(D(L),L).
\end{equation}
Our goal is to compute numerically the l.h.s. of Eq.~(\ref{RG_general}), without any \emph{apriori} assumption about the single-parameter scaling. The single-parameter scaling implies that the $\beta$-function depends only on $D(L)$ and thus the solution $L(D)$ of this equation
is a single-valued function. The inverse function $D(L)$ may be few-valued, but in any case it should be represented by a single parametric curve. On the contrary, if there are other (hidden, or irrelevant in the RG language) parameters, there will be a family of curves satisfying Eq.~(\ref{RG_general}), each one corresponding to a specific initial condition. Thus numerical evaluation of the l.h.s. of Eq.~(\ref{RG_general}) provides a framework for  understanding the nature of the transition and the accuracy of one-parameter scaling.

Before discussing the numerical results, we would like to review the general properties of the $\beta$-function if the single-parameter scaling is given for granted.

\subsection*{General properties of \texorpdfstring{$\beta(D)$}{beta(D)}}  
In the localized phase, when $D\ll D_c$, and in particular when $D\to 0$, the eigenfunctions decay exponentially with the distance from the localization center $\psi^2(r)=A\,r^{-\alpha}\,\exp[-r/\xi]$. Moreover, in a finite spatial dimension $d$, the number of sites at a given distance $r$ grows as $ n(r) \sim r^{d-1}$. Therefore, the participation entropy becomes 
\begin{align}
\label{eq:S_exp}
    S &\equiv -\Big \langle\sum_{x}\psi^{2}(x)\,\ln \psi^{2}(x) \Big \rangle \nonumber \\
    &\simeq - \sum_{r=0}^L n(r) A r^{-\alpha} e^{-r/\xi} \,\ln( A r^{-\alpha} e^{-r/\xi}) \nonumber \\
    &= - \ln A + \sum_{r=0}^L n(r) A r^{-\alpha} \left( \alpha \ln r +\frac{r}{\xi} \right) e^{-r/\xi},    
\end{align}
with $\sum_{r=0}^L n(r) A r^{-\alpha} e^{-r/\xi} = 1$ from the wavefunction normalization. From the definition of $D(L)$, ~\eqref{D_L_def}, and neglecting $\ln r$ in~\eqref{eq:S_exp}, being subleading, we get  
\begin{equation}
    D \simeq \frac{L^{2-\alpha} n(L)}{\xi d} A e^{-L/\xi} \simeq \left( \frac{L}{\xi}\right)^{d+1-\alpha}\,e^{-L/\xi}, 
\end{equation}
and using this result, the $\beta$-function turns out to be
\begin{equation}
\label{eq:beta_D0}
    \beta(D)\approx \frac{1}{d}\ln D -\frac{d-\alpha+1}{d}\ln|\ln D|, \;\;\;D_{c}\gg D \gtrsim 0.
\end{equation}
At not very large $d\sim 1$ and $\alpha\sim 1$ the first term makes the leading contribution to $\beta(D)$ in the insulator.  In the limit $d\rightarrow\infty$ we have $D_{c}\rightarrow 0$ and 
  the region of applicability of Eq.~(\ref{eq:beta_D0}) shrinks to zero. This is a clear indication of the failure of single-parameter scaling to describe this limit properly.

In the other limiting case $D\rightarrow 1$ we have $\beta(D)\simeq \alpha_1 (1-D)$. The slope $\alpha_1$ can be easily found from the results of Ref. ~\cite{abrahams1979scaling}. Close to the metallic limit in the orthogonal ensemble, the corrections to $D$ must be proportional to the inverse of the dimensionless conductance:
\begin{equation}
    D \approx  1-c/g \simeq 1-c'/L^{d-2}.
\end{equation}
This means that 
\begin{equation}
    \beta_d(D) = \frac{d-2}{d} \frac{1-D}{D}, \qquad D\lesssim 1,
\end{equation}
which gives
\begin{equation}
\label{eq:alpha_1}
    \alpha_1=\frac{d-2}{d}
\end{equation}
near $D=1$.
In the limit $d\to\infty$ one obtains    $\alpha_1=1$, the same scaling we find in RMT and for expander graphs \cite{sierant2023universality,vanoni2023renormalization}.  At $d=2$ we obtain  $\alpha_1 = 0$; we investigate more in detail the consequences of this observation later.  Notice also that the above result obtained using a scaling argument can also be found performing the $\epsilon$-expansion around $d=2$, as shown later.
\\

At the Anderson localization transition (and in general close to  an unstable  fixed point of the RG equations) we must have 
\begin{equation}\label{beta_crit}
    \beta(D)=\alpha_c (D-D_c),
\end{equation}
where we are assuming that $\beta(D)$ vanishes with a finite derivative; such assumption is valid in any finite dimension but is not necessarily true in the $d\to \infty$ limit~\cite{vanoni2023renormalization}.  Later on we argue that, for short-range models like the Anderson model on a $d$-dimensional lattice, $\alpha_c$ remains finite in this limit.

The slope $\alpha_{c}$ determines the finite-size scaling exponent $\nu$. Indeed, plugging Eq.~(\ref{beta_crit}) into Eq.~(\ref{RG_general}) and setting $D\approx D_{c}$ one finds the solution: 
\begin{equation}
    \ln|D-D_c|-\ln|D_{0}-D_{c}|=\alpha_c D_c d\ln L,
\end{equation}
where $D_{0}$ is the value of $D(L)$ at the smallest $L\sim 1$.
Then one readily obtains:
\begin{equation}\label{L_cr}
    D=D_c\pm (L/\xi)^{1/\nu},\;\;\;\;\xi\sim |D_{0}-D_{c}|^{-\nu}\sim |W_{0}-W_{c}|^{-\nu}.
\end{equation}
where $\nu$ is the finite-size scaling exponent:
\begin{equation}
\label{eq:nu-alpha-Dc}
    \nu=1/(\alpha_c d D_c).
\end{equation}
The values of $\alpha_c$ and $D_c$ depend on $d$ and must be found from the numerics. 
\section*{$\beta(D)$ in the $\epsilon$-expansion and self-consistent theories}

In this Section, we discuss the properties of the single-parameter $\beta(D)$ function using the  previous analytical results obtained within the sigma-model formalism and certain self-consistent theories of localization.  

\subsection*{$\epsilon$-expansion within non-linear sigma-model}
\label{sec:epsilon_exp}
 Here we employ the results of Refs.\cite{Wegner1987, Wegner1989} in $d=2+\epsilon$ dimensions which are obtained from the formalism of the nonlinear sigma model. In the orthogonal symmetry class, they read:
\begin{equation}\label{beta-t}
-\frac{d\ln t}{d\ln L^{d}}=\frac{\epsilon}{d} - \frac{2}{d}\, t- \frac{12 \zeta(3)}{d}\, t^{4}+O(t^{5}),
\end{equation} 
\begin{equation}\label{D-eps}
1-D_{q}^{(c)}=\frac{q \epsilon}{d}  + \frac{\zeta(3)}{4d}\, q (q^{2}-q+1) \epsilon^{4}+O(\epsilon^{5}),
\end{equation}
where $t(L)$ is the inverse dimensionless conductance, using the same notation as in the literature, and $D_{q}^{(c)}$ is the $q$-th fractal dimension at $W=W_c$ (for the definition of $D_{q}^{(c)}$ see Ref.\cite{evers2008anderson}).

Now we introduce the scale-dependent fractal dimension $D_{q}(L)$ away from the criticality and find the corresponding $\beta$-function. To this end we use the single-parameter scaling that implies $D_{q}(L)=D_{q}(t(L))$ and require that $D_{q}(t^{*})=D_{q}^{(c)}$ given by Eq.~(\ref{D-eps}), where $t^{*}$ is the fixed point of RG equation Eq.~(\ref{beta-t}).

Then expressing $\epsilon$ in terms of $t^{*}$ from Eq.~(\ref{beta-t}), plugging it in Eq.~(\ref{D-eps}) and replacing $t^{*}$ by $t=t(L)$ we obtain for $D(L)\equiv D_{q=1}(t(L))$:
\begin{equation}\label{D-t}
1-D=(2/d)(t+8\zeta(3)\,t^{4}).
\end{equation}
Differentiating Eq.~(\ref{D-t}) with respect to $L^{d}$, using the RG Eq.~(\ref{beta-t}), expanding in $t\ll 1$ up to $t^{4}$ and using Eq.~(\ref{D-t}) we finally obtain:
\begin{align}\label{beta-fin}
\beta(D)=&(1-D)\,\left(1-\frac{2}{d D} \right)+3 d^{2}(d-2)\zeta(3)\,(1-D)^{4}\\ \nonumber \
&-\frac{d^{2}(24-d)}{4} \zeta(3)\,(1-D)^{5}+O[(1-D)^{6}].
\end{align}
At small $d-2=\epsilon$ one can expand Eq.~(\ref{beta-fin}) up to quadratic order in $(1-D)$:
\begin{equation}
\label{eq:beta_eps_d2}
\beta_{D}(D)=(\epsilon/2)(1-D) - (1-D)^2+O((1-D)^3).
\end{equation} 
Notice that the coefficient 1 of $(1-D)^2$ agrees with what is extracted from the numerical data (see Fig.~\ref{fig:beta_2} and Eq.~(\ref{eq:beta_d2})). This is an independent check of our numerical procedure.
In this parabolic approximation the slopes of the $\beta$-function, $\alpha_{c}$ and $\alpha_{1}$,  at $D=D_{c}$ (where $\beta_{D}(D_{c})=0$) and $D=1$,   obey the symmetry:
\begin{equation}\label{sym}
\alpha_{c}=-\alpha_{1}=\frac{\epsilon}{2}.
\end{equation}
However, this symmetry breaks down even in the $\epsilon^{2}$ approximation when the subtle terms $\sim \zeta(3)$ are still neglected.
\subsection*{The self-consistent theory by Vollhard and Woelfle and its violation}
\begin{figure}[t]
    \centering
    \includegraphics[width=0.45 \textwidth]{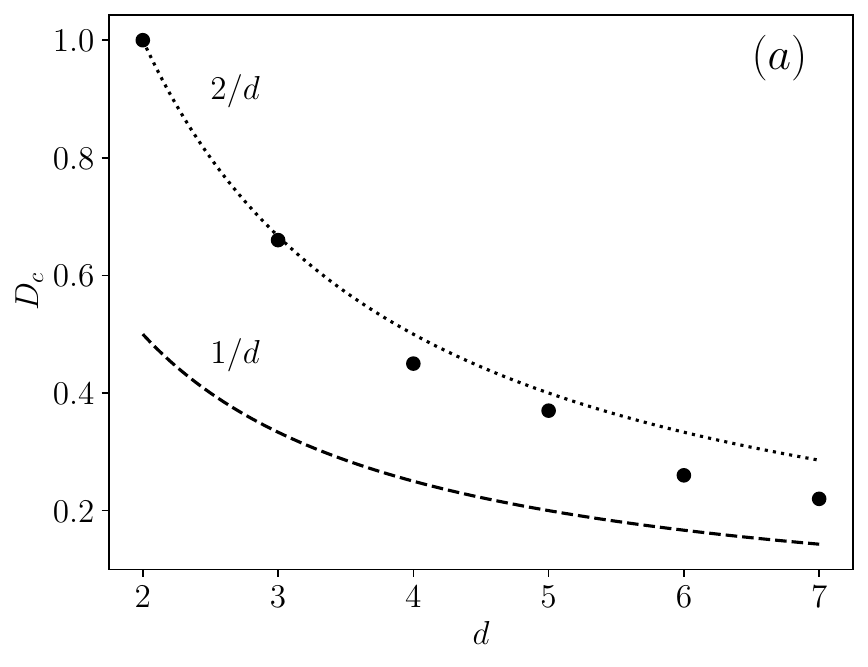}
    \includegraphics[width=0.45 \textwidth]{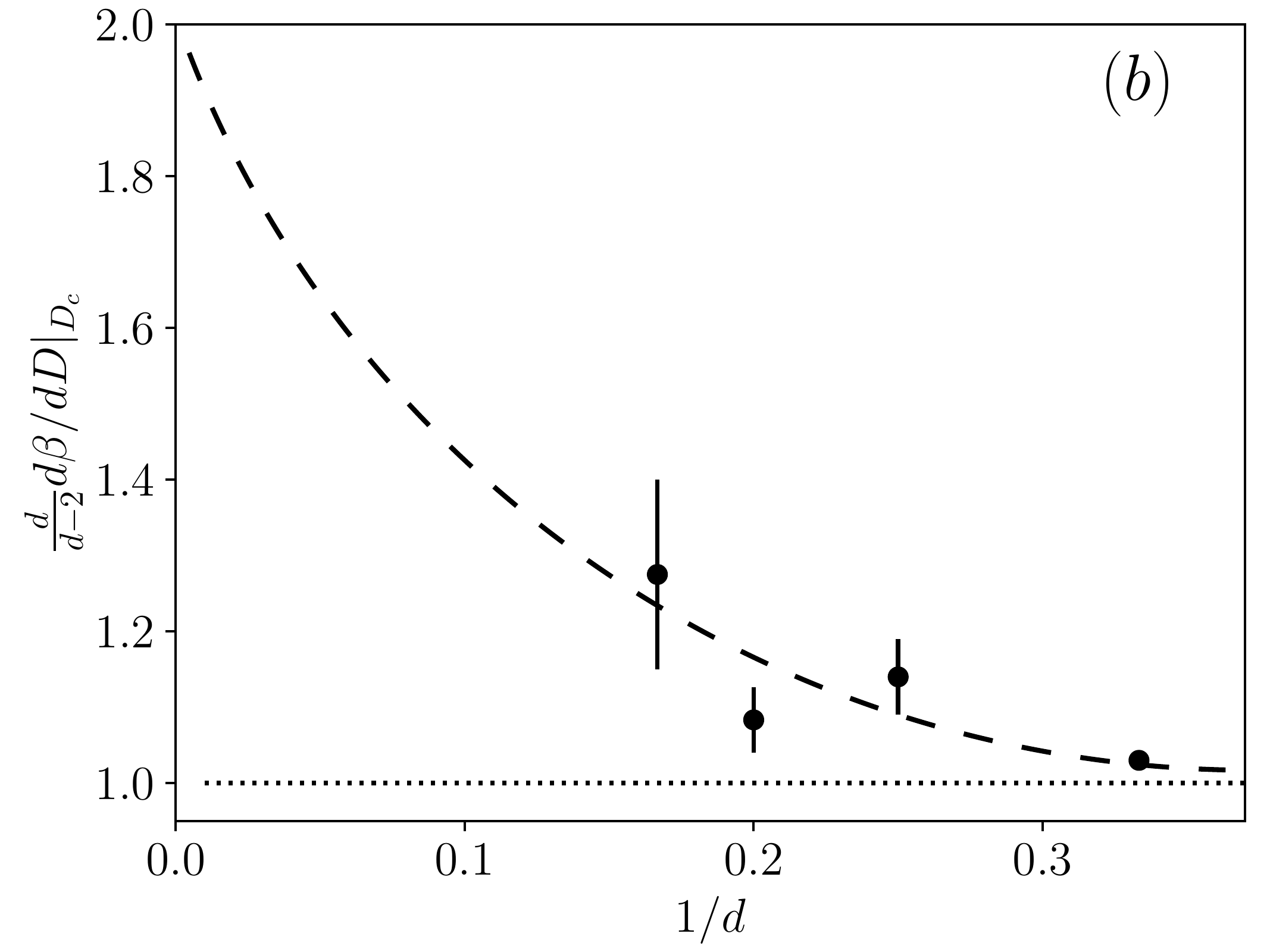}
    \caption{(a) The critical fractal dimension $D_{c}$  and  (b) the ratio of slopes $d\beta_{D}/dD$ at $D=D_{c}$ and $D=1$ as a function of $d$ from exact diagonalization of the Anderson model on $d$-dimensional lattice. The last point $d=7$ on the left panel is obtained with very restricted system sizes $L<7$ and by a simplified method different from all other points. The dashed line in panel (b) qualitatively illustrates our conjecture, Eqs.~(\ref{bound}),(\ref{alpha-conj}). }
	\label{fig:bounds}
\end{figure}
In the absence of the four-loop corrections proportional to $\zeta(3)$ the critical point $D_{c}$, the slope $\alpha_{c}$ of the $\beta$-function at $D=D_{c}$ and the critical exponent $\nu$, Eq.~(\ref{eq:nu-alpha-Dc}), are found from Eq.~(\ref{beta-fin}) as:
\begin{eqnarray}
 \label{Dc-VW} D_{c}&=&\frac{2}{d},  \\
\label{alpha_c-VW} \alpha_{c}&=&\frac{d-2}{2}, \\
 \label{nu-VW}\nu&=&\frac{1}{d-2}.
\end{eqnarray} 
This result coincides with the one of the so-called “self-consistent theory of localization" by Vollhardt and Woelfle (VW)\cite{wolfle2010self}. If Eqs.~(\ref{Dc-VW}),(\ref{alpha_c-VW}),(\ref{nu-VW}) were exact for some $d<d_{\mathrm{up}}$, where $d_{\mathrm{up}}$ is an upper critical dimension, then inevitably $d_{\mathrm{up}}=4$, as for $d=4$ the exponent $\nu$ takes its mean field value $\nu=1/2$. Furthermore, at $d=4$ within the VW theory, we obtain $D_{c}=1/2$ which is the lower limit for $D$ where two randomly chosen fractal wave functions intersect and thus can be correlated resulting in the Chalker's scaling \cite{Chalker_Daniel, Chalker}.  

As a matter of fact, the values of $D_{c}$ at $d=3$ and $d=4$  match Eq.~(\ref{Dc-VW}) pretty well (but $D_c$ is smaller than $2/d$ for $d>4$, see Table I). However,
the value of $\nu\approx 1.57-1.59$ at $d=3$, that is found numerically in Refs.\cite{Slevin-PRL99,ohtsuki1999review, Roemer2011, ueoka2014dimensional}, differs substantially from the result of this theory $\nu=1.0$, thus invalidating it. Therefore, there is no reason to trust the result of the VW theory, Eqs.~(\ref{alpha_c-VW}),(\ref{nu-VW}), according to which the slope $\alpha_{c}$ diverges in the limit $d\rightarrow\infty$. 

Also, the values of $\alpha_{c}$ which we found and collected in Table I, are not described by Eq.~(\ref{alpha_c-VW}). Surprisingly, for $d=3,4$ the values of $\alpha_{c}$ are very close to $-\alpha_{1}$ thus approximately exhibiting the symmetry Eq.~(\ref{sym}) which should hold only at small $d-2=\epsilon$.

In fact, the contribution of the higher-order terms in the loop expansion in the nonlinear $\sigma$-model (the second and the third term in Eq.~(\ref{beta-fin})) makes the critical $D_{c}$ smaller than $2/d$: 
\begin{eqnarray} 
\label{Dc-expan}  D_{c}=\frac{2}{d}- \frac{\zeta(3)}{8} \,\epsilon^{4}+O(\epsilon^{5}),
\end{eqnarray}
The slope $\alpha_{c}$ and the ratio $\alpha_{c}/\alpha_{1}$ is also affected by these terms:
\begin{eqnarray}
\label{alpha_c-expan}&&\alpha_{c}\equiv \left(\frac{d\beta_{D}}{dD}\right)_{D=D_{c}}= \frac{\epsilon}{2}+\frac{9}{8} \zeta(3)\,\epsilon^{4}+O(\epsilon^{5}),\\
\label{alpha_1}&&\alpha_{1}\equiv\left(\frac{d\beta_{D}}{dD}\right)_{D=1}=- \frac{d-2}{d},\\
\label{ratio} &&\left|\frac{\alpha_{c}}{\alpha_{1}}\right|=1+\frac{\epsilon}{2}+\frac{9}{4}\zeta(3)\,\epsilon^{3}.
\end{eqnarray}
Notice that the product $\alpha_{c}D_{c}$ that determines the exponent $\nu$ reproduces the well-known result \cite{hikami1992localization} obtained from the $\beta$-function for the variable $t$, Eq.~(\ref{beta-t}):
\begin{equation}
\label{one_over_nu}  \alpha_{c}D_{c}d = \nu^{-1}=\epsilon + \frac{9}{4}\zeta(3)\,\epsilon^{4}+O(\epsilon^{5}). 
\end{equation}	
This demonstrates the invariance of $\nu$ with respect to the change of variables $t(L)\rightarrow D(L)$ and provides a proof of the correctness of our perturbative calculations.
 
Despite Eqs.~(\ref{Dc-expan})-(\ref{alpha_c-expan}) and Eqs.~(\ref{ratio})-(\ref{one_over_nu}) are valid only at very small $\epsilon\lesssim 0.1$ and do not apply even for the case $d=3$, the tendency they show is correct and observed in the numerical simulations [see Fig.~\ref{fig:bounds}(a,b)]. In particular, the fact that $D_{c}$ decreases faster than $2/d$ with increasing $d$ and that the ratio of the slopes obeys the following inequality:
\begin{equation}\label{bounds}
\left|\frac{\alpha_{c}}{\alpha_{1}}\right|>1,
\end{equation}
and grows with increasing $d$, is convincingly confirmed. 
\subsection*{Correlation between $D_{c}$ and $\alpha_{c}$ and a ‘semi-classical theory' for $\nu_{d}$}
As was already mentioned, the critical $D_{c}$ for $d=3,4$ is very close to the result of the VW self-consistent theory $D_{c}=2/d$. Next, we would like to note that the derivative of the $\beta$-function $\alpha_{1}=-(d-2)/d$ at the fixed point $D=1$ is an exact result of Eq.~(\ref{beta-fin}) which is independent of the higher-order terms in $(1-D)$. It is interesting to see what would happen if $D_c = 2/d$ and the symmetry Eq.~(\ref{sym}) is enforced beyond the lowest $\epsilon$-expansion. The immediate consequence of $\alpha_c = |\alpha_1|$ is that the exponent $\nu = (d D_c \alpha_c)^{-1} = (d D_c |\alpha_1|)^{-1}$ would take the form:
\begin{equation}
\label{eq:nu-semi-cl}
\nu=\frac{d}{2(d-2)}=\frac{1}{2}+\frac{1}{d-2}.
\end{equation}
Surprisingly, we obtained the formula empirically suggested by many authors \cite{Garcia-Garcia, Garcia-Garcia_Cuevas, ueoka2014dimensional}, most notably in Ref.~\cite{Garcia-Garcia} where a sort of derivation is presented in the spirit of VW self-consistent theory. We think, however, that this ‘semiclassical theory' is seriously flawed. In this derivation the momentum dependence of a Cooperon was changed from $\xi^{2} q^{2}$ in the original VW paper to $ D_{0}\,\xi^{2} q^{d}$, while the dependence of the correlation length $\xi$ remained the same. This inevitably requires the dependence of $D_{0}\propto \ell^{d-2}$ on the ultraviolet cutoff $\ell^{-1}$ which violates the single-parameter scaling. In contrast, our numerics demonstrates that the single-parameter scaling at $d=3,4,5,6$ is a very reasonable approximation. 

Notwithstanding this comment, the values of $\nu$ obtained from Eq.~(\ref{eq:nu-alpha-Dc}) for $d=3,4,5,6$ using $D_{c}$ and $\alpha_{c}$ found directly from the single-parameter $\beta$-function (see Fig.~\ref{fig:beta_d}), are very close to the ones following from Eq.~(\ref{eq:nu-semi-cl}), obtained numerically in Ref.~\cite{Garcia-Garcia_Cuevas} and also experimentally in Ref.~\cite{madani2024exploring} for $d=4$. At the same time, the values of $D_{c}$ and $\alpha_{c}$ significantly differ from $2/d$ and $(d-2)/d$, respectively (see Fig.~\ref{fig:bounds}). This implies highly correlated deviations of these quantities from the above naive predictions.

We would like to stress that, in order to obtain a single-parameter curve, we employed a procedure that is completely different from the numerical approach of Refs.\cite{Slevin-PRL99,Roemer2011,ohtsuki1999review,ueoka2014dimensional,slevin2014}. In our approach, we extracted the single-parameter curve with no assumption on the number and values of the irrelevant exponents and then determined the relevant exponent $\nu$ from this single-parameter curve. This procedure is more complicated compared to that of Refs.\cite{Slevin-PRL99,Roemer2011,ohtsuki1999review,ueoka2014dimensional,slevin2014} and it inevitably leads to less accurate numerical estimates of the exponents \footnote{Since the procedure involves finding the maximum of the numerical $\beta(D,L)$ in a given small interval $[D,D+\Delta D]$, for different $L,W$, we believe our procedure can lead to a {\it systematical overestimate} by a few percent the values of $\alpha_c, D_c$ and hence of $\nu$.}. 
However, the clear advantage of this procedure is that it gives a detailed picture of the RG flow and emergence of single-parameter scaling and it is free from the choice of the number and values of the irrelevant exponents. In any case, the surprisingly high accuracy of a simple formula Eq.~(\ref{eq:nu-semi-cl}) for different dimensionalities $d=3,4,5,6$ raises again a question of its status and the approximation (which we think is still lacking) it can be obtained from. 
\section*{A conjecture about the lower bound on $D_{c}$}
In the absence of the finite upper critical dimension $d_{\mathrm{up}}$ (i.e. $d_{\mathrm{up}}=\infty$) it seems plausible that the exponent $\nu$ tends to 1/2 in the limit $d\rightarrow\infty$, as was suggested by a number of authors (see e.g. Ref. \cite{ueoka2014dimensional}).
Then the slope $\alpha_{c}$ in this limit can be found from~\eqref{eq:nu-alpha-Dc} as:
\begin{equation}\label{alpha-inf-d}
\alpha_{c}=\frac{2}{D_{c}\,d}.
\end{equation}
An immediate consequence of this is that $\alpha_{c}$ is finite in the $d\rightarrow\infty$ limit if $D_{c}$ decreases with increasing the dimensionality $d$ as $D_{c}\propto 1/d$ and this slope has an infinite limit if $D_{c}$ decreases faster than $1/d$.
Unfortunately, the numerical data up to $d=7$ of Table 1 and Fig.~\ref{fig:bounds} allows both asymptotic behaviors, with a crossover dimensionality that we estimate around $d^{*}\sim 10$. In this situation of the lack of theory at large $d$ (when the non-linear sigma model is no longer justified) and the inability of numerical simulation on the lattices of dimensionality $d\gg d^{*}$ we would like to propose a conjecture on the lower bound for $D_{c}$ for the Anderson model on $d$-dimensional lattices with short-range hopping.
We argue that 
\begin{equation}\label{bound}
D_{c}\geq \frac{1}{d}.
\end{equation}
and if the upper critical dimension $d^{\mathrm{up}}=\infty$ this inequality saturates only at $d=\infty$. 

The reason for this conjecture is that by definition $D_{c}=d_{c}/d$, where $d_{c}$ is the dimensionality of the support set of multifractal wave function embedded into a lattice of dimensionality $d$. Clearly, if $d_{c}<1$ the support set cannot be connected and should look like a set of points with the typical distance $R$ between them much greater than the lattice constant.  For a lattice model with short-range hopping, at high dimensions $d$ the critical disorder $W_{c}\sim d \ln d$ is large. Therefore the typical transmission amplitude between such points is exponentially small  $G\propto {\exp[-R \ln(W_{c})]}$. The points may belong to the same support set only if their on-site energies are in resonance with an exponential accuracy, i.e. $\Delta E\lesssim {\exp[-R \ln(W_{c})]}$. This situation is exponentially rare, as the probability of a resonance is $\sim \Delta E/W_{c}$, and this is exactly the point why we believe $d_{c}$ must be greater than 1 if the wave function is extended and the model is short-ranged. 

Certainly, this argument does not apply to systems with long-range hopping, e.g. for the Power-Law Banded random matrices  \cite{PLBRM} or the Rosenzweig-Porter models \cite{RP15,LN-RP20,kutlin2024investigating}. In those cases, $d=1$ and it is known that $d_{c}<1$ can be arbitrarily small.

If the conjecture Eq.~(\ref{bound}) is true then Eq.~(\ref{alpha-inf-d}) immediately gives:
\begin{equation}\label{alpha-conj}
\lim_{d\rightarrow\infty}\alpha_{c}=2,
\end{equation}
that is, it is (a) finite and (b) twice larger than $\lim_{d\rightarrow\infty}\alpha_{1}=1$. 
This seemingly innocuous conclusion has an important implication for the critical scaling of the Anderson model on Random Regular Graphs (RRG). If, in fact, $\beta(D)_{\mathrm{RRG}}=\lim_{d\to\infty}\beta(D)_d$, then this allows us to choose scenario I formulated in Ref.~\cite{vanoni2023renormalization} as the only possible, and therefore the RRG has two diverging lengths as $W\to W_c$: one with exponent $\nu=1/2$ and one with exponent $\nu=1$, which dominates (although sizes larger than the available ones are needed to observe $\nu=1$ in the numerical data). The existence of two critical exponents was also discussed, in a different context, in Ref.~\cite{Mata2020two}.

\section*{Numerical $\beta$-function}
\subsection*{Numerical $\beta$-function in two dimensions}
\begin{figure}[t]
    \centering
    \includegraphics[width=\linewidth]{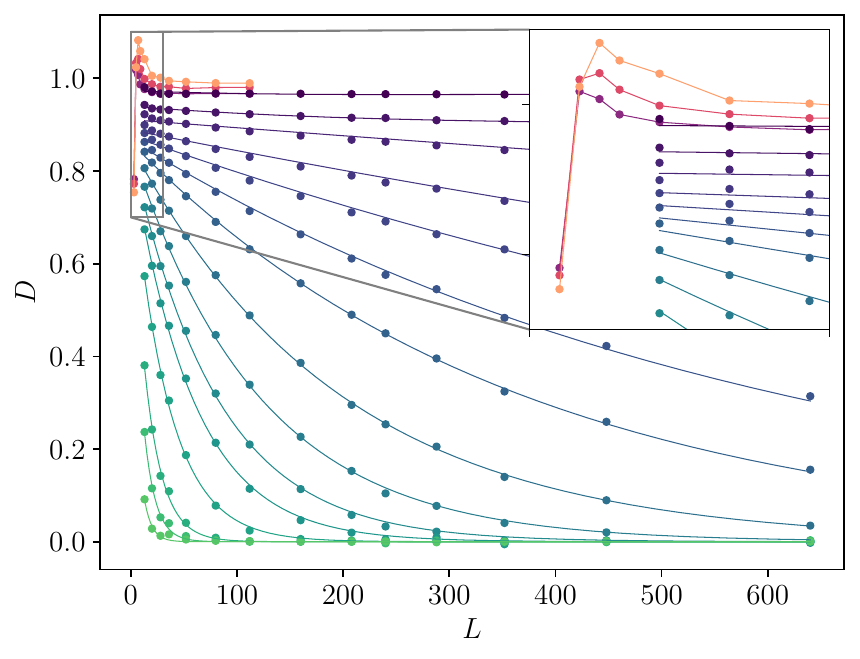}
    \caption{System size dependence of $D(L)$ in $d=2$, for different values of $W$. The solid lines in shades of green are interpolations of the data, used to produce the $\beta$-function in Fig.~\ref{fig:beta_2}. For small sizes and small $W$, $D(L)$ may exceed $D=1$ even if the system is localized in the thermodynamic limit, as shown in the inset by the red-shaded curves. This behavior can be obtained analytically if the eigenfunction inside the localization radius is weakly multifractal. It happens because of the ‘basin' regions where the eigenfunction amplitude $\psi^{2}\sim N^{-1-\eta}$ decreases with the volume faster than $N^{-1}$. Such regions should have a large enough probability to overcome the dominance of the ergodic regions with $\psi^{2}\sim N^{-1}$ in the normalization sum $\sum_{r}\psi(r)^{2}=1$. The job to suppress the probability of ergodic regions is done by the regions with ‘elevated' $\psi^{2}\sim N^{-1+\eta}$ which are always present in a weakly multifractal state together with the ‘basin' areas. A similar behavior of $D(L)$ is present in the Rosenzweig-Porter random matrix ensemble \cite{RP15, LN-RP20}.  }
    \label{fig:d2_D}
\end{figure}

\begin{figure}[t]
    \centering
    \includegraphics[width= \linewidth]{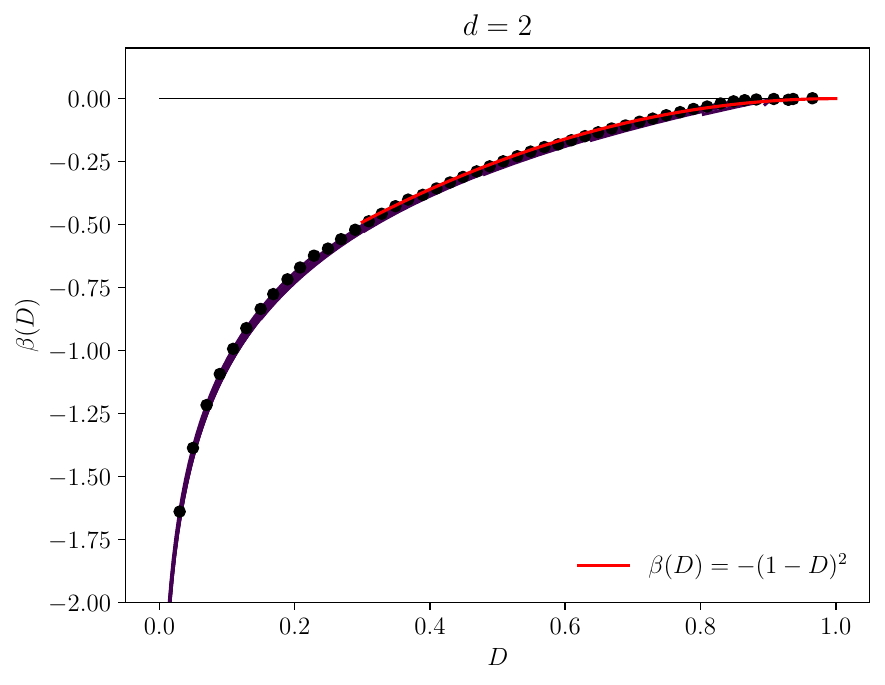}
    \caption{Plot of the $\beta(D)$ for the Anderson localization model on a two-dimensional  lattice. The dark lines are the numerical results, that are obtained from the participation entropy according to the definition, and the black dots are the envelope of the data, identifying the one-parameter scaling part of the $\beta$-function. In particular, the fractal dimension is computed by applying the discrete derivative to $S$, and the resulting points are interpolated using a Padé fit for $W<W_c$ and an exponential fit for $W>W_c$ (see Fig.~\ref{fig:frac_dim_L} for more details on the interpolations). The red curve is $\beta(D) = - (1-D)^2$, which perfectly fits the data and coincides with the correction given by the sigma model, according to Eq.~(\ref{eq:beta_eps_d2})}
    \label{fig:beta_2}
\end{figure}

Now we present the numerical results on the $L$-dependence of $D(L)$ and obtain the $\beta$-function for a two-dimensional system.  
The set of data on the $L$-dependence of $D(L)$ for different $W$ obtained from Eq.~(\ref{D_L_def}) is presented in Fig.~\ref{fig:d2_D}, where the Shannon entropy is computed from Eq.~(\ref{eq:vN_entr}) using the eigenfunctions from the exact diagonalization of the Anderson model and averaging over disorder and eigenfunctions. From this set of data we obtain the plot $\beta(D)$ vs $D$ which is presented in Fig.~\ref{fig:beta_2}.
Remarkably, all the RG trajectories lie almost exactly on a single curve, just corroborating the single-parameter scaling as a very precise approximation in $d=2$. More details about the $\beta$ function in $d=2$ can be found in the SI Appendix.
\subsection*{Numerical $\beta$-function in three dimensions} 
 \begin{figure}[t]
    \centering
    \includegraphics[width= \linewidth]{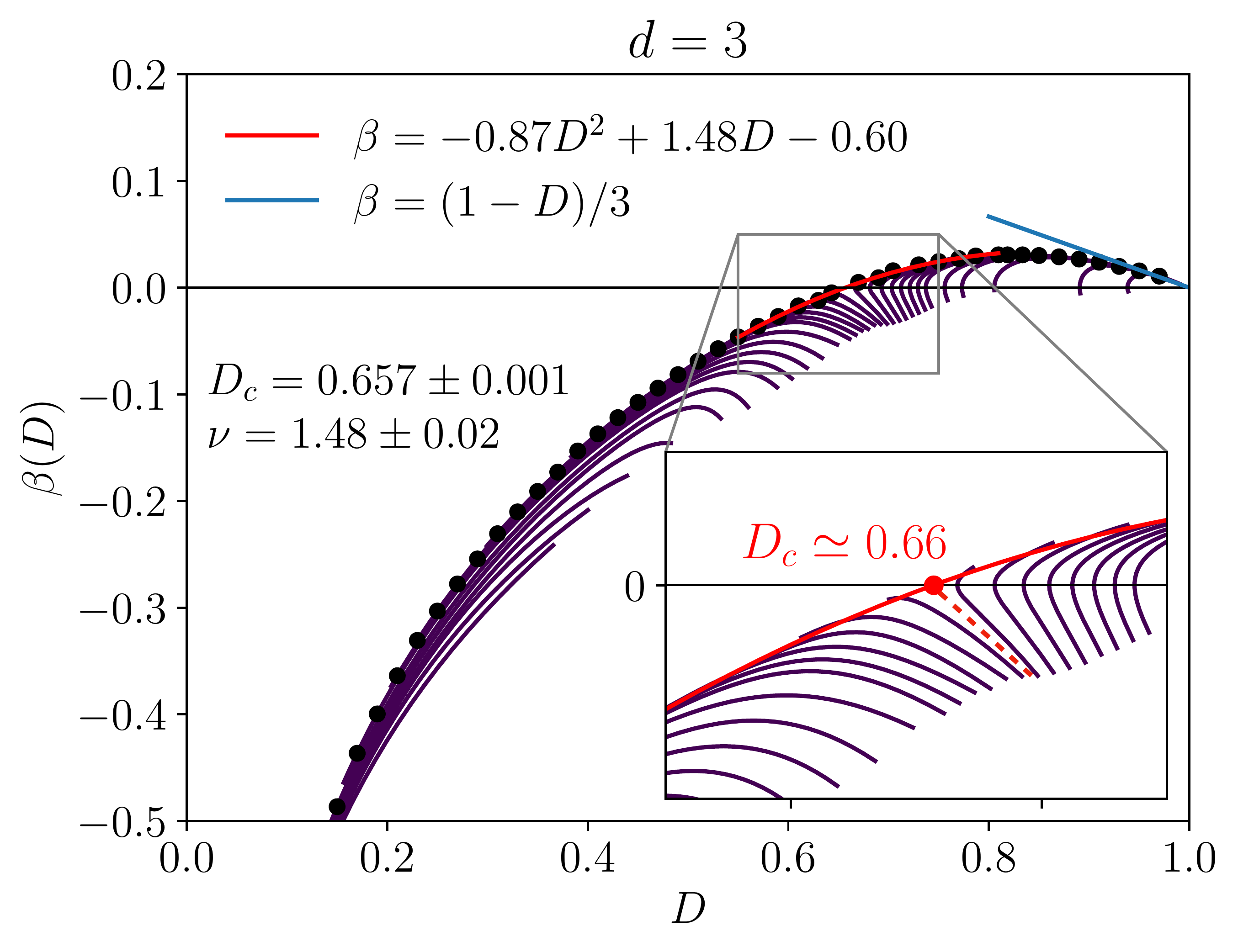}
    \caption{Renormalization group (RG) trajectories (solid lines) for $3d$ Anderson model obtained from the numerical calculation of the eigenfunction Shannon entropy $S(L)$ and the corresponding finite-size fractal dimension $D(L)=dS(L)/d \ln N$. The envelope of RG trajectories (black dots) is the single-parameter $\beta$-function $\beta(D)$. Its root $D_{c}$ gives the fractal dimension of the critical wave functions and the slope of the red solid curve at $D=D_{c}$ determines the relevant critical exponent $\nu$.
    The slope of $\beta(D)$ at the ergodic fixed point $D=1$ (blue solid line) is $(d-2)/d=1/3$.
    The accuracy of one-parameter scaling can be inferred from the length of the initial parts of the trajectories, ‘the hairs', before merging with the single-parameter curve.}
    \label{fig:beta_3}
\end{figure}
The numerical $\beta$-function in three dimensions obtained by the same procedure is presented in Fig.\ref{fig:beta_3}. In contrast to $d=2$
the trajectories corresponding to this $\beta$-function  are  initial condition   specific  (initial conditions are $L=L_{0}$ and a certain $W$). At small enough $L$ they deviate significantly from the single-parameter curve forming a sort of “hairs" on the plot. With increasing $L$ they merge with the single-parameter curve which is a common envelope (shown by a solid red line on the plot) of all such trajectories. There is one special trajectory that terminates exactly at the fixed point (shown by a red point), so that we have a {\it critical trajectory} and not just a critical fixed point in three dimensions. The existence of “hairs" is a "smoking gun" of violation of the single-parameter scaling due to the irrelevant operators which are beyond the sigma-model description.

\subsection*{Numerical $\beta$-function for higher dimensions}
\begin{figure}[t]
    \centering
    \includegraphics[width=0.49 \linewidth]{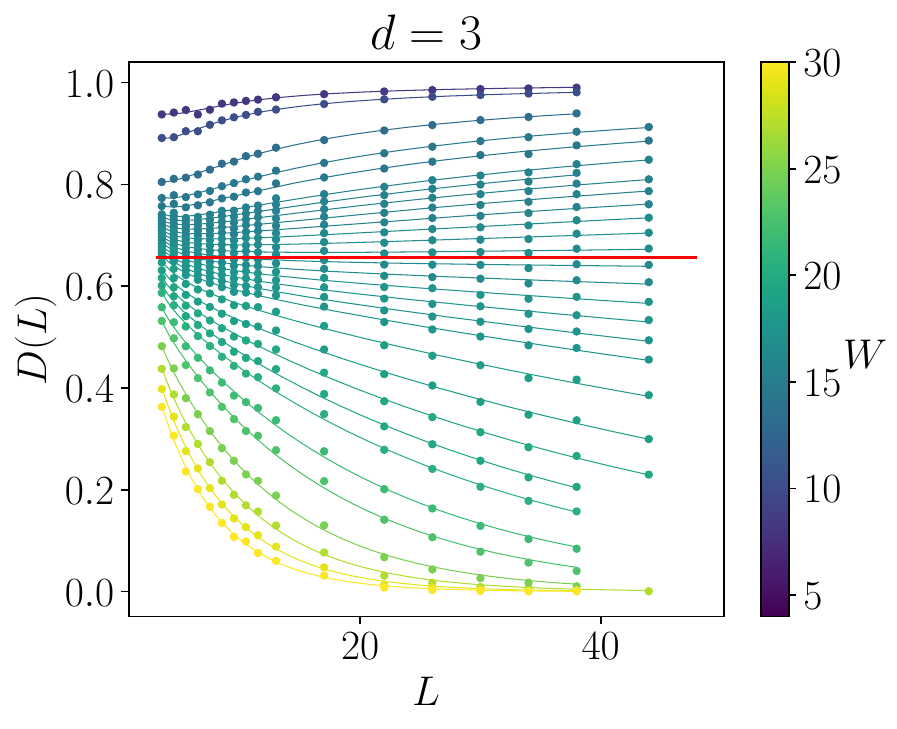}
    \includegraphics[width=0.49 \linewidth]{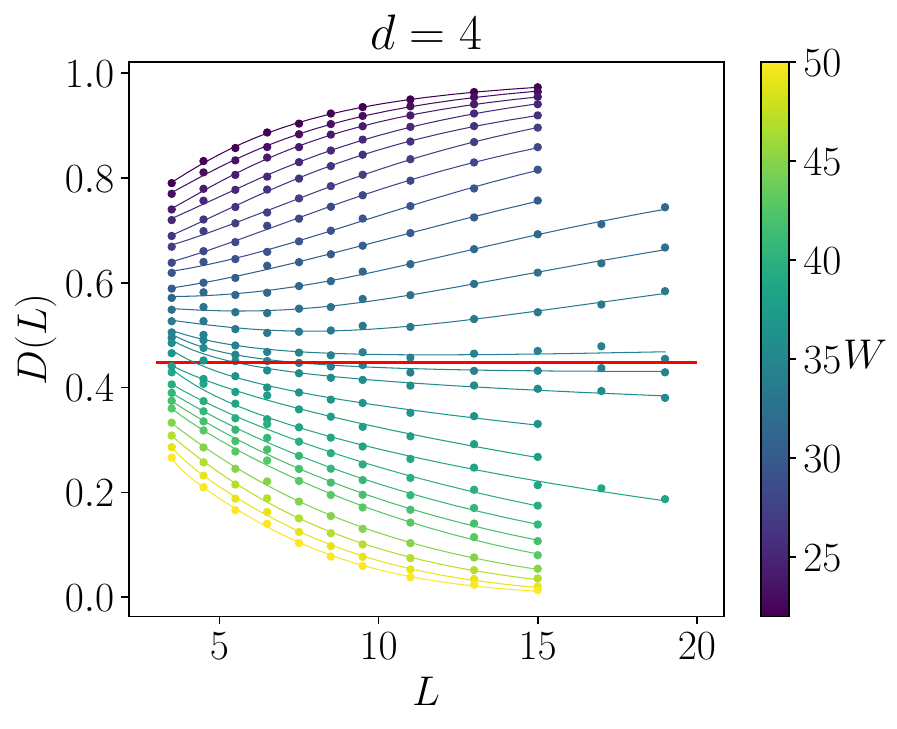}
    \includegraphics[width=0.49 \linewidth]{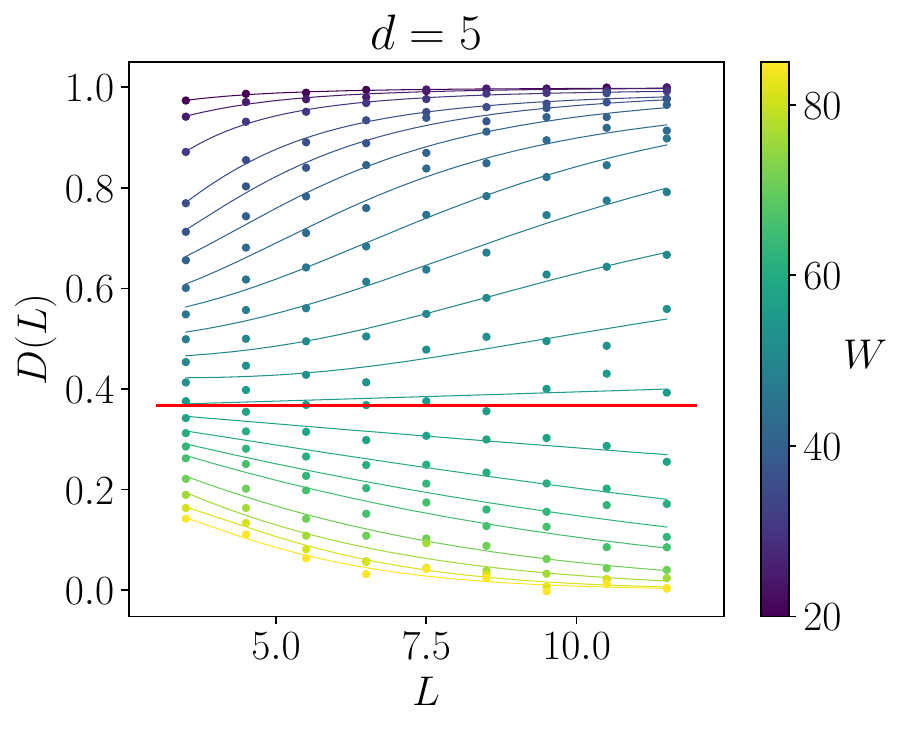}
    \includegraphics[width=0.49 \linewidth]{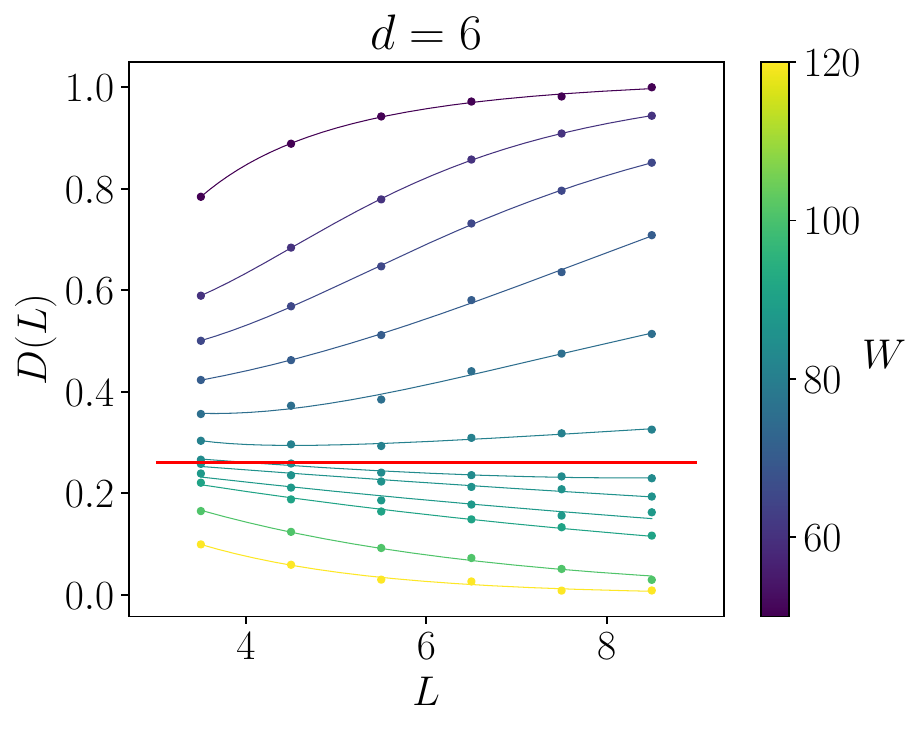}
    \caption{System size dependence of the numerical fractal dimension at different dimensions $d=3,4,5,6$ and for different values of $W$. The solid lines are \emph{interpolations} of the data, that we will use to produce the $\beta$-function. In particular, for $W<W_c$ we interpolate using a Padè function $D(L,d) = (L^{d-1} + a L^{d-2} + b)/(L^{d-1} + c L^{d-2} + k)$, while for $W>W_c$ we use $D(L) = \exp{(-aL)} (b L + c)/(k L + m)$. These choices are dictated by physical arguments, namely the behavior of $\beta(D)$ at $D\sim 1$ and the exponential decay of $D$ in the localized phase.
    The red lines in each plot represent the values of the critical fractal dimension obtained as the point at which the $\beta$-function vanishes, and that are reported in Table 1.}
    \label{fig:frac_dim_L}
\end{figure}

\begin{figure}[t]
    \centering
    \includegraphics[width=0.49 \linewidth]{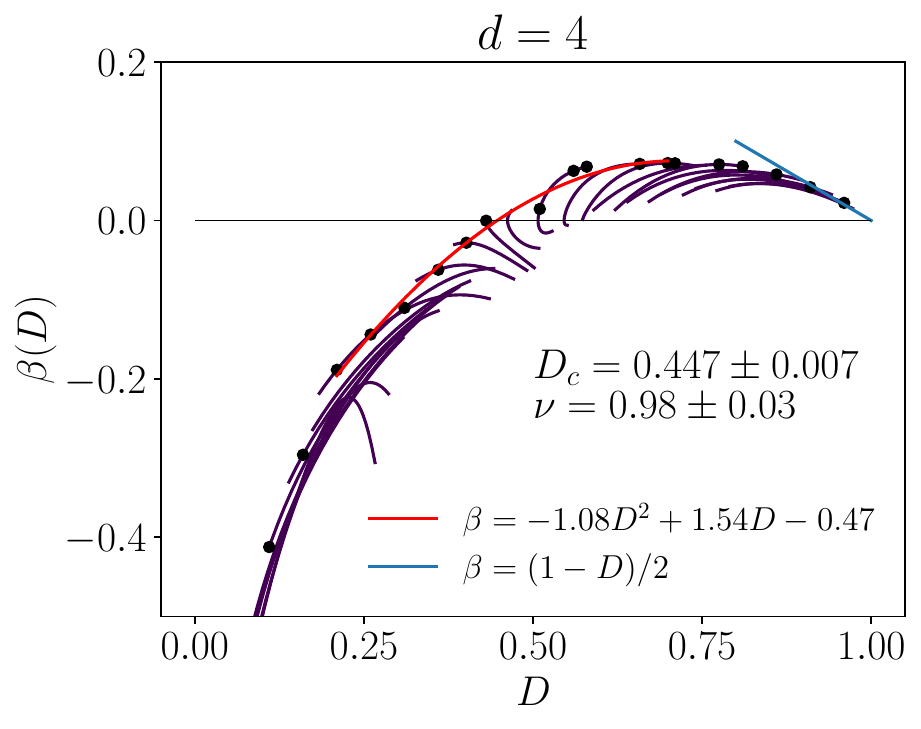}
    \includegraphics[width=0.49 \linewidth]{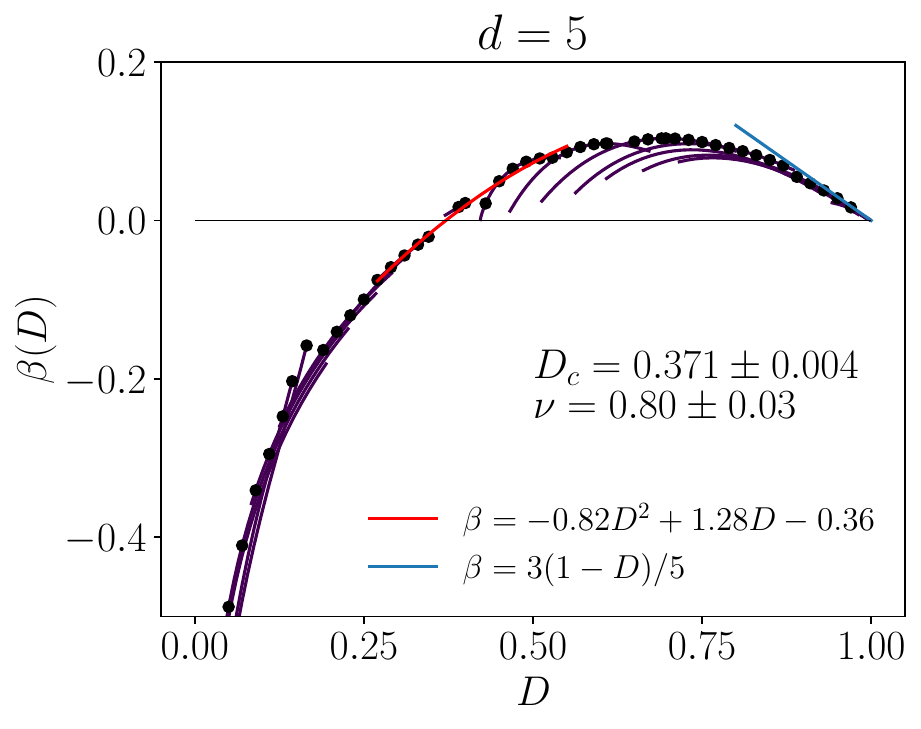}
    \includegraphics[width=0.49 \linewidth]{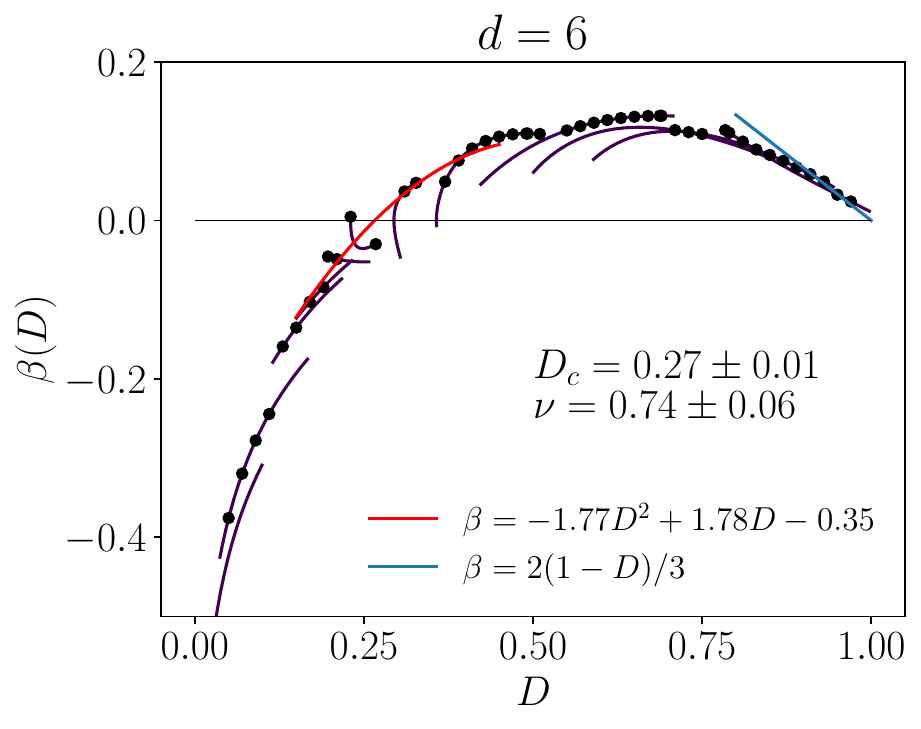}
    \includegraphics[width=0.49 \linewidth]{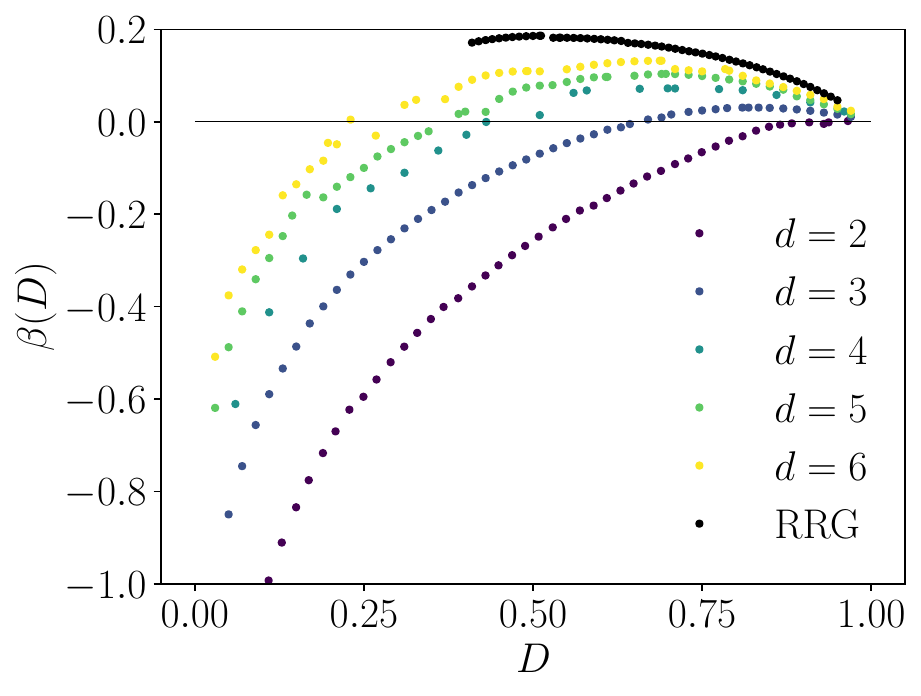}
    \caption{Plots of the $\beta(D)$ for the Anderson localization model on  higher-dimensional ($d=4,5,6$) lattices. The colored lines are the numerical results, that are obtained from the participation entropy according to the definition. In particular, the fractal dimension is computed by applying the discrete derivative to $S$, and the resulting points are interpolated using a Padé fit for $W<W_c$ and an exponential fit for $W>W_c$ (see Fig.~\ref{fig:frac_dim_L} for more details on the interpolations). The black points are a proxy for the envelope of the data, while the red curves are quadratic fits of the envelope of the numerical data around $\beta=0$, from which we extract $D_c$ and $\alpha_c$ as reported in the plots and Table 1. The blue line instead is the theoretical prediction for $\beta$ near $D=1$. The last plot is the set of envelopes for different dimensions and the RRG, displayed to highlight the flow for $d\to \infty$.}
    \label{fig:beta_d}
\end{figure}
\begin{figure}[h]
    \centering
    \includegraphics[width=\linewidth]{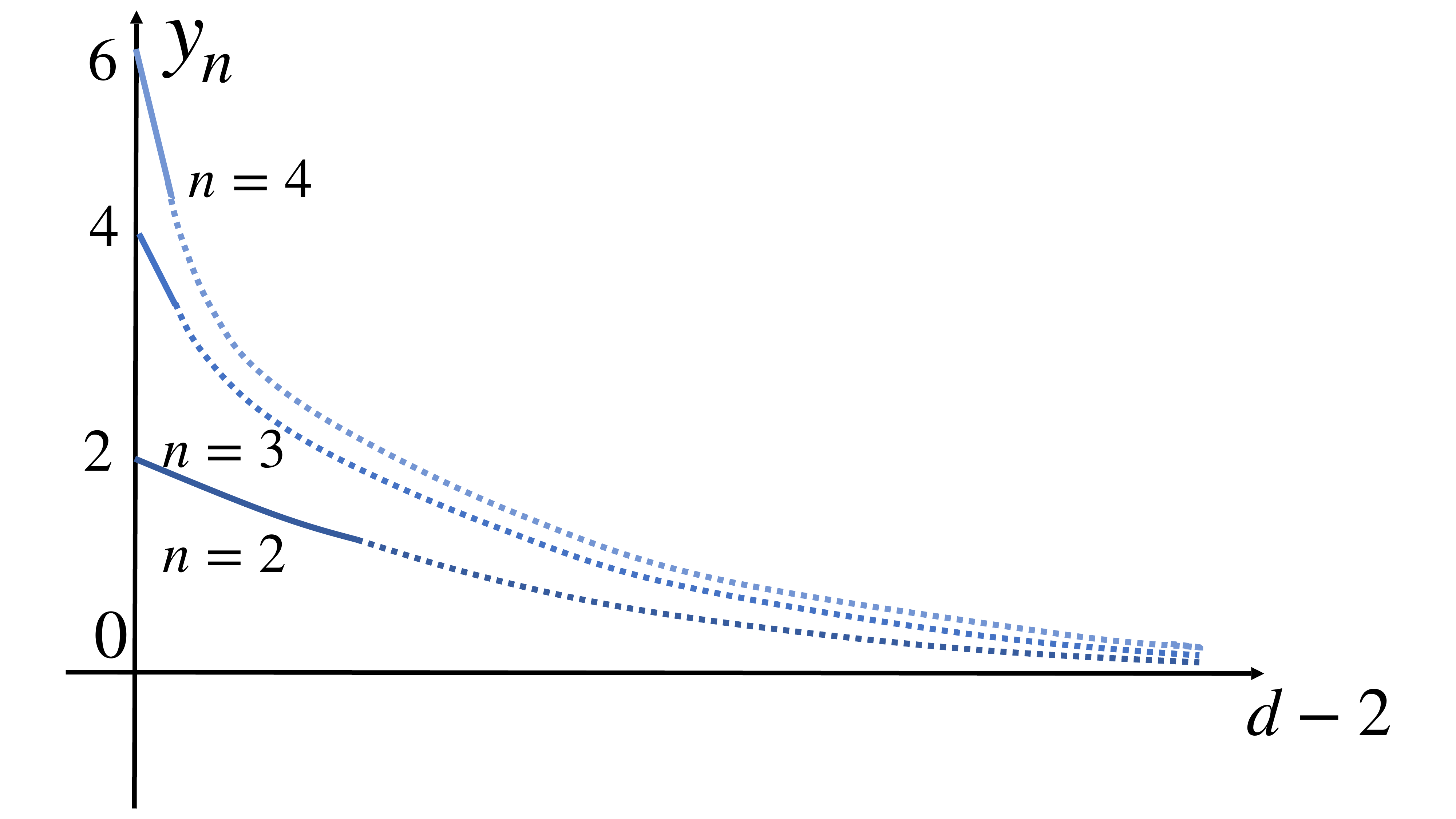}
    \caption{Dependence of the irrelevant exponents $y_n$ on the spatial dimension of the system.  The solid lines represent Eq.~(\ref{yn}), while the dashed lines give a sketch of our conjecture as $d$ increases. We conjecture that in the limit of large $d$ all dashed lines merge and approach zero.}
    \label{fig:y_n}
\end{figure}
 The $\beta$-function can be computed for higher dimensions, albeit with an accuracy that decreases as $d$ increases. The results are presented in Fig.~\ref{fig:frac_dim_L} and Fig.~\ref{fig:beta_d}.

Analyzing the results, we were able to extract the parameters $D_{c}$, $\alpha_{c}$ and $\nu$ of a single-parameter curve and compare them with the available numerical results for $\nu$ in Table 1. Surprisingly, the value of $\nu$ for $d=3,4,5,6$ extracted from the best fit of a single-parameter curve $\beta(D)$ close to critical point $D=D_{c}$ is very close to that described by the ‘semiclassical self-consistent theory' $\nu=1/2 + 1/(d-2)$, albeit the theory itself is seriously flawed.

\begin{figure}[t]
    \centering
    \includegraphics[width=\linewidth]{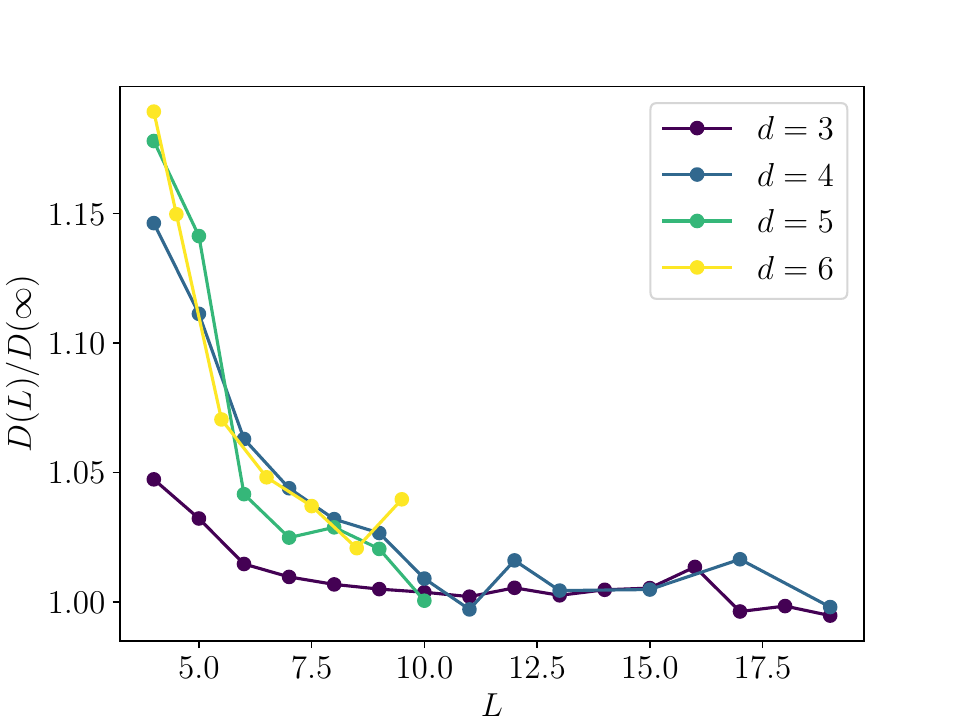}
    \caption{System size dependence of $D(L)/D(L \to \infty)$ at $W = W_c$ for different spatial dimensions. When $d$ increases, also $D(L=O(1))/D(L \to \infty)$ grows, that implies a longer length of the ‘‘hair'' in the $\beta$-function. However, the saturation value is achieved approximately at the same linear size $L=O(10)$, as we discuss in the main text.}    \label{fig:Dc_d}
\end{figure}

Another important result of our numerics is that the effect of the irrelevant exponent (encoded in the length of the RG trajectory at $W=W_{c}$ before it hits the fixed point) increases as $d$ increases (see also Fig.~\ref{fig:Dc_d}).

\section*{The high-gradient operators in the non-linear sigma-model and the irrelevant exponent $y$}
  To describe the initial part of RG trajectories (“hairs") one needs to take account of one or more  {\it irrelevant exponents} $y$ introduced in Ref.~\cite{Slevin-PRL99}. Apparently, such exponents are {\it beyond} the single-parameter scaling as described by the formalism of the conventional non-linear sigma model \cite{wegner1979mobility, Wegner1987, Wegner1989}. 

In order to understand the origin of the operators corresponding to the exponent $y$ one has to {\it extend} the conventional sigma-model \cite{wegner1979mobility,Efetov_sigma-model}. The corresponding extension was done in Ref.~\cite{AKL_Lett, AKL_JETP, AKL_book} by adding to the sigma-model, in addition to the conventional ‘diffusion' term $t^{-1}\,{\rm Str}[(\nabla Q)^{2}]$, also  the higher-order ($n>1$) terms of the gradient expansion:
\begin{equation}\label{high-grad}
Z_{n}\,\ell^{2(n-1)}\,{\rm Str}[(\nabla Q)^{2n}],
\end{equation}  
where $Q$ is the Efetov's super-matrix \cite{Efetov_sigma-model}, $\ell$ is the electron mean free path and ${\rm Str}$ denotes the super-trace. Such terms can be rigorously derived \cite{AKL_Lett, AKL_JETP, AKL_book} starting from the model of free electrons in impure metals. 

The additional terms have an irrelevant exponent $-y_{n}^{(0)}=-2(n-1)$ in the zero-order approximation of non-interacting diffusion modes (the conventional term proportional to $(\nabla Q)^{2}$ has an exponent $0$ in this approximation). The interaction of diffusion modes leads to a renormalization of the coupling constant $t$ described by one-parameter scaling, Eq.~(\ref{beta-t}). However, it also gives rise~\cite{AKL_Lett, AKL_JETP, AKL_book}  to renormalization of $Z_{n}$ in Eq.~(\ref{high-grad}):
\begin{equation}\label{RG-Z}
\frac{d \ln Z_{n}}{d u}= n(n-1)+ \mathrm{higher\; order\; in\;} t\sim \epsilon,
\end{equation}
where 
\begin{equation}\label{u}
u=\ln\left( \frac{\sigma_{0}}{\sigma(L)}\right)=\frac{(L/\ell)^{\epsilon}}{1+(L/\xi)^{\epsilon}}.
\end{equation}
Here $\epsilon=d-2$, $\xi$ is the critical length, $\sigma_{0}$ is the Drude conductivity, and $\sigma(L)$ is that with effects of localization included.

At small $\epsilon$ one may neglect the higher-order terms in $t\sim \epsilon$ in Eq.~(\ref{RG-Z}), so that:
\begin{equation}\label{Z}
Z_{n}=Z_{n}^{(0)}\, \left[ \frac{(L/\ell)^{\epsilon}}{1+(L/\xi)^{\epsilon}}\right]^{n(n-1)}.
\end{equation}
At criticality, $L \ll \xi$, the $L$-dependent term in the denominator of Eq.~(\ref{Z}) can be neglected and we obtain $Z_{n}\propto L^{\epsilon\,n (n-1)}$. This gives a {\it positive} correction to $y_{n}$ (see also Fig.~\ref{fig:y_n}):
\begin{equation}\label{yn}
y_{n}=2(n-1)- \epsilon\, n(n-1) +  o(\epsilon).
\end{equation}
 At $\epsilon\ll 1$ the largest irrelevant exponent corresponds to $n=2$, so that we obtain:
\begin{equation}\label{y}
y=y_{2}=2-2\epsilon + o(\epsilon).
\end{equation}
Equation~(\ref{y}) shows that the irrelevant exponent $y<2$ (which is always the case in numerics \cite{Roemer2011}) and decreases with increasing the dimensionality.  As usual in $2+\epsilon$-expansion in the localization problem, this equation is not applicable already for $d=3$. However, it shows a tendency towards making the irrelevant exponent less irrelevant with increasing $d$. This results in the corrections to single-parameter scaling   more significant (and hence the length of “hairs" longer, 
see Fig.~\ref{fig:beta_d}), as $d$ increases.

What happens at large $d$? One of the possibilities is that the irrelevant exponent becomes relevant (positive) at some finite $d=d^{\mathrm{up}}$ and the single-parameter scaling will no longer hold for $d>d^{\mathrm{up}}$, even as an approximation. We, however, think that $d^{\mathrm{up}}=\infty$ and the breakdown of the single-parameter scaling happens only for localization problems on expander graphs like Random Regular Graph (RRG). Notice that according to  Eq.(\ref{yn})  the initial slope $n(n-1)$ in the dependence of $y_{n}$ on dimensionality grows faster with $n$ than its value $2(n-1)$ at $d=2$ increases. This makes it possible that all $y_{n}$ merge and accumulate  near zero  in the limit $d\rightarrow\infty$ (see dotted lines in Fig.\ref{fig:y_n}). Such an accumulation could be a mechanism of a complete breakdown of single-parameter scaling on Random Regular Graph \cite{vanoni2023renormalization}.

We would like to emphasize that the scenario of breakdown of single parameter scaling at $d>d^{\mathrm{up}}$ described above is different from the one suggested recently by Zirnbauer \cite{arenz2023wegner, Zirnbauer23}. The true theory of the NEE phase with singular-continuous spectrum should, perhaps, be a combination of both, in which the higher-gradient terms should play an important role.
\begin{table*}[t]
\centering
\begin{tabular}{c|c|c|c|c|c|c}
    $d$ & $W_c$& $D_c$ & $\alpha_c d$ & $\nu=\frac{1}{\alpha_c d D_c}$ & $\nu_{{\rm num}}$  & $\nu$, Eq.~(\ref{eq:nu-semi-cl})  \\
    \hline
    $3$ & $16.4 \pm 0.2$ & $0.657 \pm 0.001$ & $1.029 \pm 0.01$ & $1.48 \pm 0.02$ & $1.57\pm 0.004$ \cite{slevin2014}, $1.52\pm 0.06$ \cite{Garcia-Garcia_Cuevas}  & $3/2$  \\
    \hline 
    $4$ & $34.3 \pm 0.2$ & $0.447 \pm 0.007$ & $2.28\pm 0.10$  & $0.98 \pm 0.03$ & $1.156\pm 0.014$   \cite{ueoka2014dimensional}, $1.03\pm 0.07$ \cite{Garcia-Garcia_Cuevas} & $1$ \\
    \hline
    $5$ & $56.5 \pm 0.5$ & $0.367 \pm 0.004$ & $3.25 \pm 0.13$ & $0.84 \pm 0.03$ & $0.969\pm 0.015$ \cite{ueoka2014dimensional}, $0.84\pm 0.06$ \cite{Garcia-Garcia_Cuevas} & $5/6\approx 0.83$ \\
    \hline
    $6$ & $83.5 \pm 0.5$ & $0.26 \pm 0.01$ & $5.1 \pm 0.5$ & $0.74 \pm 0.06$ & $0.78\pm 0.06$ \cite{Garcia-Garcia_Cuevas} & $3/4$ \\
    \hline
    $7$ & $110 \pm 2$ & $0.22 \pm 0.04$ & / & / & / & / \\
\end{tabular}
\label{table:exponents}
\caption{Numerical values for critical properties in $d=3,4,5,6$, compared with previous results in the literature. The values of $W_c$ we find, corresponding to the red lines in Fig.~\ref{fig:frac_dim_L}, are compatible with the results in the literature~\cite{slevin2014,Tarquini2017critical}. The values of $D_c$ and critical exponents are found by analyzing the numerical data around $\beta=0$. The errors displayed are the ones coming from a quadratic fit of the envelope of the $\beta$-function near the critical point (red curve in the plots). We expect the actual errors to be larger than the ones reported.}
\end{table*} 
\section*{Approaching the critical point}
In the vicinity of the critical point, $\beta$ stops being a single-valued function of $D$ since the contributions of irrelevant parameters become important. In a previous work \cite{vanoni2023renormalization} we have seen that in the RRG this means that a square root behavior of the $\beta$-function is observed. We now show that this is also the case in finite $d$ and that this is in one-to-one correspondence with the introduction of corrections to scaling due to irrelevant parameters.

We start by the corrected scaling form, in the close vicinity of the critical point $W=W_c$ (see \cite{sierant2023universality}), defining $\tau=(W-W_c)/W_c$ and including only one irrelevant operator of dimension $-y<0$:
\begin{eqnarray}\label{scal}
        D(W,L)=f_0(L^{1/\nu}\tau)+L^{-y}f_1(L^{1/\nu}\tau).
\end{eqnarray}
The functions $f_{0,1}$ are smooth, monotonic functions of their arguments (universal scaling functions).

The localization/critical length is $\xi\sim |\tau|^{-\nu}$. The upper branch of $\beta$ describes the situation in which $L\gg \xi\sim |\tau|^{-\nu}$ and therefore the second term is negligible with respect to the first. The only zero of $\beta$ in this case is given by the single-parameter scaling behavior and we know this implies $\beta(D)\simeq (1/d D_c \nu)(D-D_c)$. However, if we take limit in which $L\ll |\tau|^{-\nu}$, we can have another zero of $\beta$, and this is the origin of the square root singularity at $D=D_A$. To find it, we simply impose
\begin{eqnarray}
    0&=&\frac{dD}{dL}=L^{1/\nu-1}\frac{1}{\nu}\tau f_0'(L^{1/\nu}\tau)\nonumber\\
    &-&y L^{-y-1}f_1(L^{1/\nu}\tau)+L^{-y}L^{1/\nu-1}\frac{1}{\nu}\tau f_1'(L^{1/\nu}\tau).
\end{eqnarray}
Now, for $L\ll |\tau|^{-\nu}$ we can neglect the third term wrt the first two, and we can also set the arguments of $f_{0,1}$ to zero. This means
\begin{eqnarray}
    0=L^{1/\nu-1}\frac{1}{\nu}\tau f_0'(0)-y L^{-y-1}f_1(0),
\end{eqnarray}
which is solved to give the length scale
\begin{equation}
    L_A^{-y-1/\nu}=\frac{\tau f_0'(0)}{y\nu f_1(0)}.
\end{equation}
This equations has a real solution only for $\epsilon f_0'(0)/f_1(0)>0$, so there is only one other zero of $\beta$ in the vicinity of $D_c$ and this is either in the localized or delocalized region depending on the behavior of the universal scaling functions $f_{0,1}$. We know that in the Anderson model, irrespective of $d$, this occurs in the delocalized region $W<W_c$.

\begin{figure}[t]
    \centering
    \includegraphics[width=\linewidth]{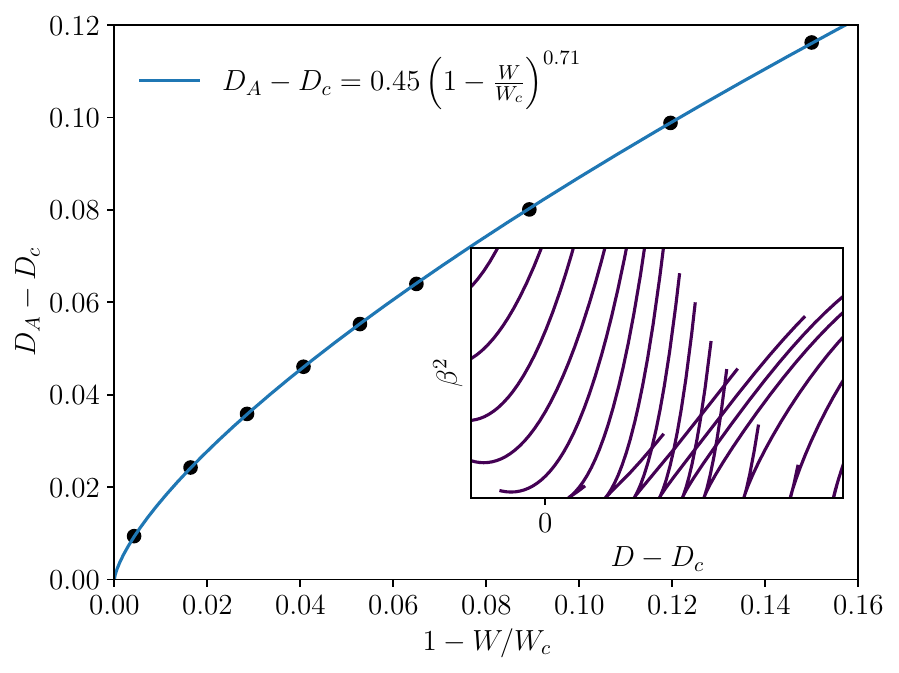}
    \caption{\emph{(Main)} Dependence of $D_A - D_c$ on $\tau = 1-W/W_c$ for $d=3$. $D_A$ is the value at which, for fixed $W$, the $\beta$-function vanishes. In producing the plot, we used $D_c = 0.657$ and $W_c = 16.47$, in agreement with the data reported in Table 1. The result of a fit of the form~\eqref{eq:fit_DA_Wc} is reported in the legend and depicted as a blue line in the plot. \emph{(Inset)} Plot of $\beta^2$ as a function of $D-D_c$: the value of $D_A - D_c$ corresponds to the cusp point of the curves. Notice that, differently from the RRG (see Fig. 10 in the Supp. Mat. of Ref.~\cite{vanoni2023renormalization}), the two branches corresponding to $\beta>0$ and $\beta<0$ soon acquire a different derivative that implies a strong asymmetry of RG trajectory near $D=D_{A}$.}
    \label{fig:DA_Wc}
\end{figure}

The new length-scale $L_A\sim |\tau|^{-\frac{\nu}{1+y\nu}}\ll \xi\sim |\tau|^{-\nu}$ and therefore defines a pre-asymptotic length-scale, which grows with a new, derived, critical exponent $\nu/(1+y\nu)<\nu$. By inserting back into Eq.(\ref{scal}) and denoting $f_0(L_{A}^{1/\nu}\tau)\approx f_{0}(0)=D_{c}$ and $f_1(L_{A}^{1/\nu}\tau)\approx f_{1}(0)=c$ we find that the position of the crossing $D_A$ (shown by a black point in Fig.\ref{fig:beta_sketch}) is 
\begin{equation}\label{DA-Dc}
    D_A=D_c+c\tau^\frac{y\nu}{y\nu+1}.
\end{equation}
We can test this prediction on the data for $d=3$, where we have abundant, high-precision numerics.
By obtaining $D_A$ from the numerical data we see a scaling (see Fig.~\ref{fig:DA_Wc})
\begin{equation}
\label{eq:fit_DA_Wc}
    D_A=D_c+c\ \tau^{0.71\pm 0.05},
\end{equation}
while using $\nu=1.58,\ y=2$ and Eq.(\ref{DA-Dc}) we have an exponent $0.759$. So the two results are perfectly compatible \footnote{The errorbar reported in~\eqref{eq:fit_DA_Wc} is only estimated, as there are many possible sources of error. The fit reported in Fig.~\ref{fig:DA_Wc} is very precise, but there are uncertainties also on $D_c$ and $W_c$ that are difficult to trace back precisely.}.

Notice that the sub-leading exponent $y$ can also be found by letting $\tau\to 0$ first and  increasing $L$, so that the sub-leading term now dominates and we have
\begin{equation}\label{crit-slope}
    \beta(D)=\frac{d\ln D}{d\ln V}=-\frac{y}{d D_c}(D-D_c).
\end{equation}
This equation describes the critical trajectory shown in red in Fig.\ref{fig:beta_3} and Fig.\ref{fig:beta_sketch}.
As in general there is no relation between $y$ and $\nu$, we see that the slope of the lower branch of the separatrix does not have to match the slope of the upper branch. Moreover, these slopes have the opposite signs. Measuring the slope of the critical trajectory, and $D_{c}$ one can obtain the subleading exponent $y$ from Eq.(\ref{crit-slope}).

As the dimension $d$ increases, the first irrelevant exponent $y$ decreases. This makes the difference between the length-scale $L_A$ and the correlation less pronounced. This has to be taken into account in analyzing the data at high dimensions $d=4,5,6,\dots$. 

\section*{Increasing space dimensionality and the Random Regular Graph}
\begin{figure}[t]
    \centering
    \includegraphics[width=\linewidth]{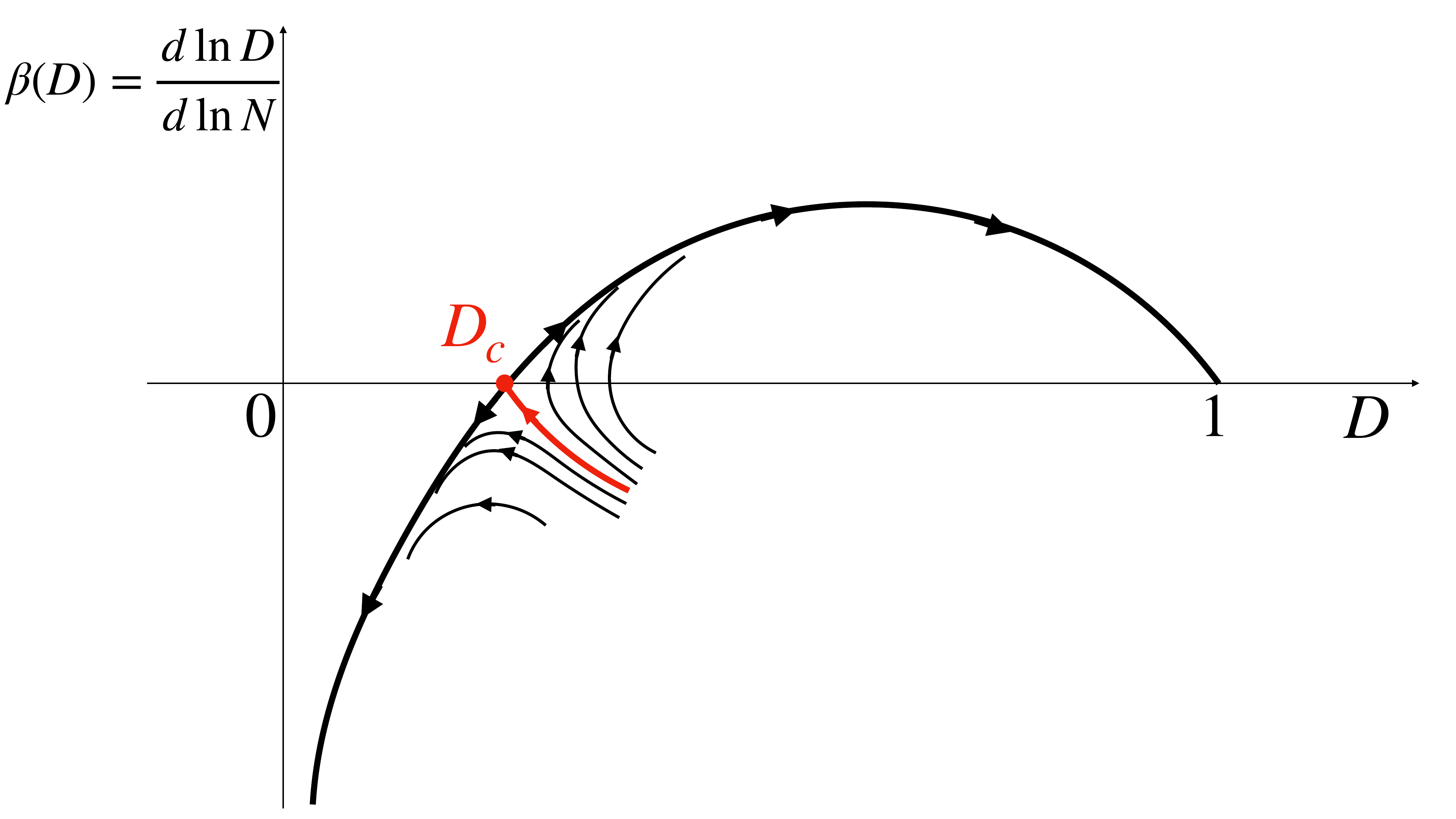}
    \includegraphics[width=\linewidth]{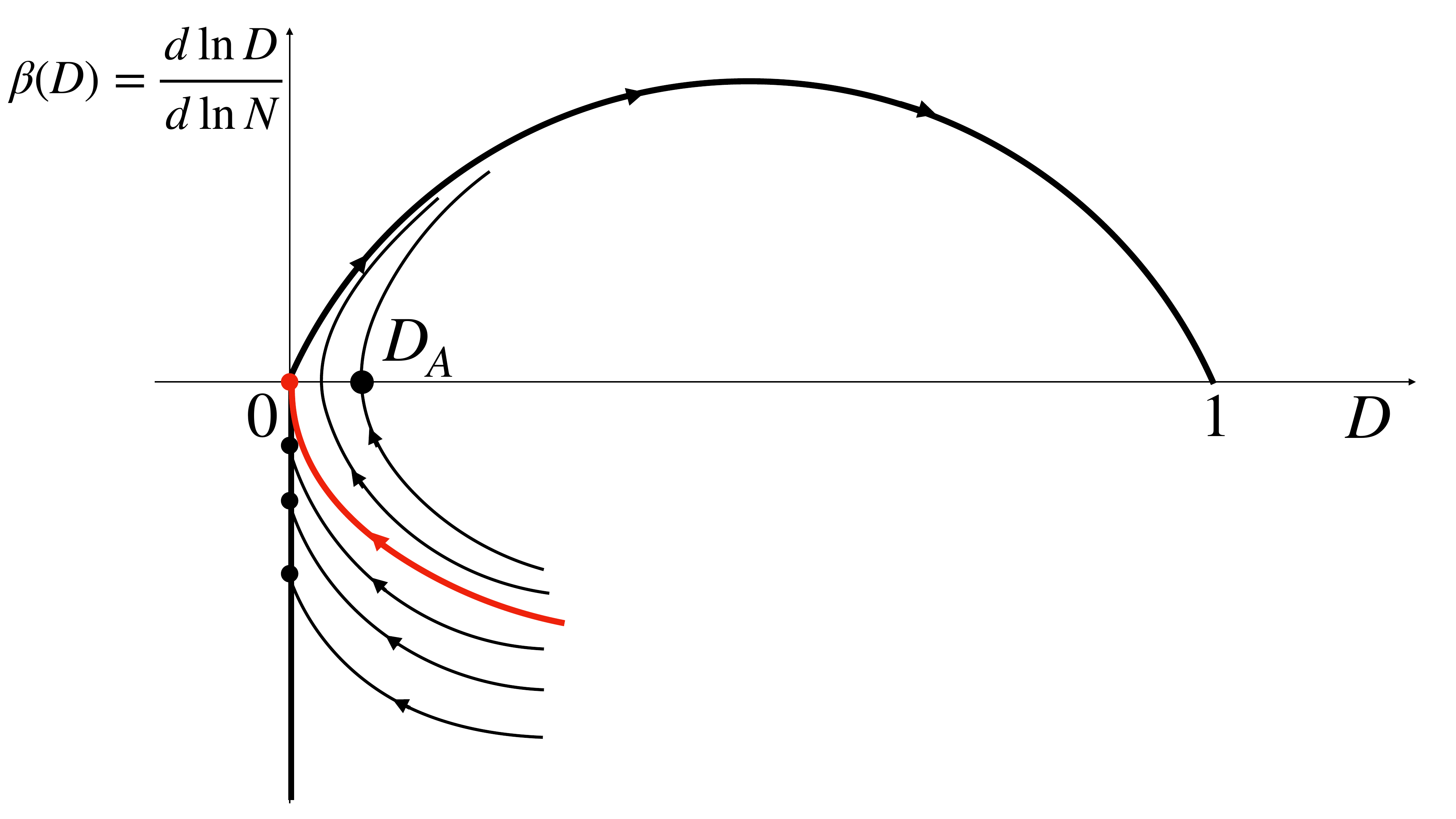}
    \caption{A sketch of the full $\beta$-function. (Upper panel) Behavior at finite dimension, where $0<D_c<1$ and the irrelevant direction at finite size becomes increasingly important as $d$ grows. (Lower panel) Behavior on expander graphs (as the RRG), where, near the critical value of $W$, the irrelevant direction becomes the only one accessible at the available system sizes.
    The critical fractal dimension $D_{c}\sim 1/d$ for finite $d$ and vanishes in the $d \to \infty$ limit, as in expander graphs. Also, the contribution of the irrelevant exponents becomes larger when $d$ grows, ultimately becoming marginal when $d\to \infty$. This is reflected by the length of the critical trajectory, depicted in red.}
    \label{fig:beta_sketch}
\end{figure}

In the previous sections, we have described in detail the behavior of the $\beta$-function for the Anderson model in finite dimensions, comparing our theoretical arguments with the numerical results from exact diagonalization.  

 The goal of this section is to summarize our knowledge and conjectures concerning the scaling behavior on a $d$-dimensional lattice in the limit $d\rightarrow\infty$.   The sketch of the evolution of the RG trajectories from a finite-$d$ lattice to the limit $d\rightarrow\infty$ represented by the Anderson model on the expander graph (e.g. on RRG) is shown in Fig.\ref{fig:beta_sketch}.

\begin{itemize}
\item
Let us first focus on the region $D \rightarrow 1$. As we already discussed, in $d$ dimensions the $\beta$-function in this limit has slope $\alpha_1 = (d-2)/d$ (see~\eqref{alpha_1}). For $d\to \infty$ this readily gives $\alpha_1 = 1$, which is the prediction of RMT and is found in the Anderson model on RRG.
\item
We have seen numerically that the critical value of the fractal dimension $D_c \leq 2/d$, and we have argued that there are reasons to believe that $D_c \geq 1/d$ for any $d$. Independently from the lower bound, $D_c \to 0$ as $d \to \infty$,  in agreement with the results on expander graphs~\cite{vanoni2023renormalization,sierant2023universality}.  
\item
 As shown in Fig.~\ref{fig:y_n} and Fig.~\ref{fig:Dc_d} and schematically sketched in Fig.~\ref{fig:beta_sketch}, the contribution of the irrelevant operators at the critical point becomes increasingly important as $d$ grows (as evident from the length of the ‘‘hairs'' in $\beta(D)$). This implies that the irrelevant exponents become less irrelevant with increasing $d$ until, eventually, a two-parameter scaling emerges for expander graphs like RRG. 
\item
There is a crucial difference between the critical length $L_{c}=(W_c -W)^{-\nu}$ when the true metallic behavior $D\approx 1$ is reached on a finite-dimensional lattice and on RRG.
For a finite-dimensional lattice the exponent $\nu=1/(d D_{c}\alpha_{c})$  found from the single-parameter scaling, depends both on the critical value of $D_{c}$ and on the slope $\alpha_{c}$ of the $\beta$-function at the critical point. In contrast, in the RRG there is always one legnthscale which has critical exponent $1/2$. In addition to this, if $\alpha_c$ is finite, there is a second length with $\nu=1$ irrespective of the value of $\alpha_c$ \cite{vanoni2023renormalization}. This crucial difference is due to $D_{c}=0$ and the fact that the critical regime is not described by the single-parameter scaling on RRG.
\end{itemize}

\section*{Role of loops and correlations in infinite dimensions}

One of the outcomes of our work is that the single-parameter  $\beta(D)>0$ (a ‘single-parameter arc') for RRG \cite{vanoni2023renormalization} is a smooth deformation of the corresponding arc for $D>D_{c}$ on $d$-dimensional lattice as $d$ increases and tends to infinity. On the other hand, it is known that in the absence of loops (i.e. on a tree)  the Anderson model (with one orbital per site) displays multifractality in the entire delocalized phase ~\cite{tikhonov2016fractality,kravtsov2018non}, where $0< D < 1$ in the thermodynamic limit. The corresponding $\beta$-function must, therefore,  terminate somewhere on the line $\beta(D)=0$ depending on the initial conditions (e.g. the strength of disorder $W$). This means that the single parameter arc in the case of a loopless tree is absent. Instead, there is a line of fixed points $[0,1]$ where the two-parameter RG trajectories terminate.  This is a strong indication that the single-parameter arc (along which the system evolves to the ergodic fixed point) emerges due to the loops on a corresponding graph.

\begin{figure}
    \centering
    \includegraphics[width=0.8\linewidth]{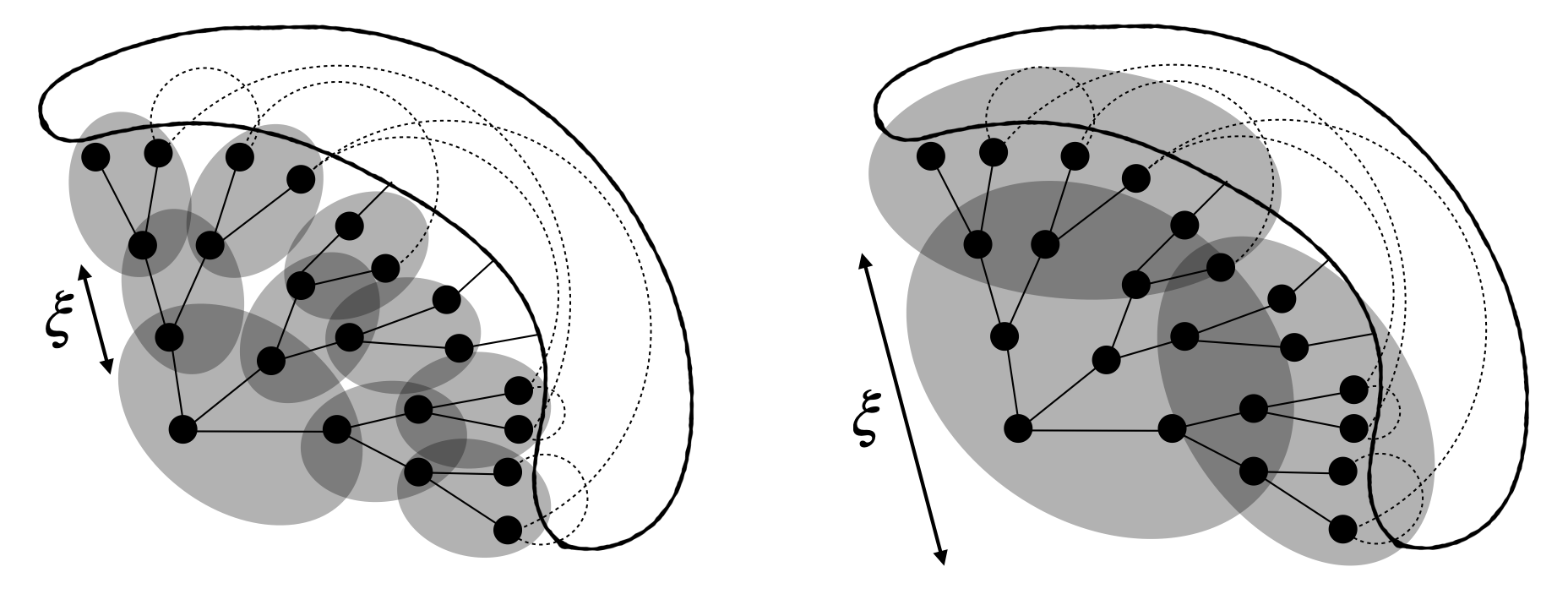}
    \caption{Pictorial representation of an RRG and the size of correlations at different disorder strengths. For small $W$ (left), the correlation length is small, and under real space RG, the limit $\xi/L \simeq 0$ is soon achieved, leading to RMT. For larger $W$, $\xi$ is larger, possibly leading to the failure of resonance hybridization, depending on the graph structure.}
    \label{fig:RRG_xi}
\end{figure}

\begin{figure}
    \centering
    \includegraphics[width=0.8\linewidth]{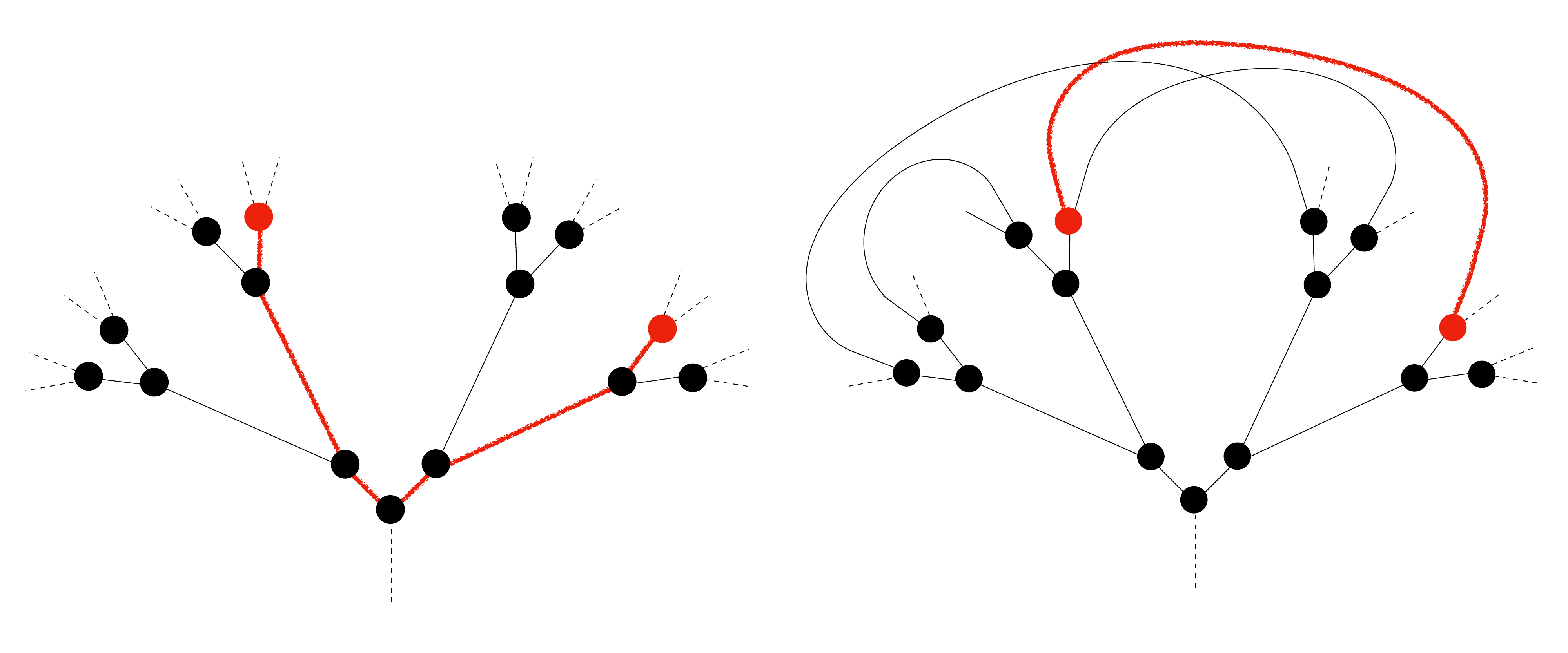}
    \caption{Resonances (pictorially represented as red dots) that are far apart on a tree (left) can become close on the RRG because of loops (right). This phenomenon facilitates the flow towards ergodicity, removing the fractal phase on the RRG.}
    \label{fig:tree_RRG}
\end{figure}

Indeed, let us consider an expander graph of diameter $L$ and connectivity $K$, so that its volume is $N = K^L$. 
In the ergodic phase, let us denote the correlation length with $\xi$, defined as the characteristic length scale for the decay of the two-point function. Upon averaging, $\xi$ is a function of the disorder strength: at small $W$, $\xi$ is small, since the system is chaotic; on the other hand, when approaching the critical point at $W = W_c$, $\xi$ diverges (see Fig.~\ref{fig:RRG_xi}), as it is expected at a phase transition. The correlation length $\xi$ can also be interpreted as the typical distance between resonances. In the localized phase the relevant length scale becomes the localization length.
When $\xi = O(1)$ resonances are very close, and under real space RG (or increasing system size) the regime $\xi/L \sim 0$ is soon achieved. The system behaves as a fully connected quantum dot and exhibits random matrix properties. By increasing $W$, the distance between resonances grows, and they can eventually fail to hybridize. Their fate, though, depends on the properties of the graph. 
On a tree, the sites on the ‘leaves' at remote branches are at a large distance from each other, as they can be connected only through the root (see right panel of Fig.~\ref{fig:tree_RRG}). 
However, RRGs are characterized by the presence of large-scale loops connecting such sites and providing the shortcuts thereby (see left panel of Fig.~\ref{fig:tree_RRG}). This means that the loops help in boosting the hybridization of resonances, reversing the RG flow and making it go to RMT along the single-parameter arc (see Fig.~\ref{fig:beta_sketch}). 

\section*{Conclusions}

In this work, we presented a renormalization group-based framework for addressing the Anderson localization transition in finite dimension. We discussed how to use the ‘modern' observables to construct the full $\beta$-function of the model in any spatial dimension $d$.  
For practical purposes, we chose the finite size fractal dimension $D(N)$ as such an observable, albeit other (eigenfunction or spectral) observables can do the same job as long as one-parameter scaling holds. We showed that some basic properties can be derived analytically by simple arguments and, when this was not possible, we presented numerical results from which we derived critical properties, in agreement with previous results in the literature. More importantly, we showed how our technique connects the perturbative results in $d=2+\epsilon$ dimensions up to $d \to \infty$, recovering the known results on RRGs, where this method has been applied recently \cite{vanoni2023renormalization}.  

We believe that the method discussed here, and already applied to expander graphs, is a new useful tool to understand the scaling properties of {\it ergodicity breaking} in disordered quantum systems, and especially to study the existence and properties of such purported transitions. This is of particular importance for interacting systems where the existence and the properties of non-ergodic phases are under long-standing debate.  

\begin{acknowledgments}

\emph{Acknowledgments.~---}

V.E.K. is grateful to Misha Feigelman for fruitful discussions and support from Google Quantum Research Award ``Ergodicity breaking in Quantum Many-Body Systems". V.E.K. is grateful to KITP, University of Santa Barbara for hospitality. This research was supported in part by the National Science Foundation under Grant No. NSF PHY-1748958.
A.S. acknowledges financial support from the National Recovery and Resilience Plan (NRRP), Mission 4 Component 2 Investment 1.3 funded by the European Union NextGenerationEU.
National Quantum Science and Technology Institute (NQSTI), PE00000023, Concession Decree No. 1564 of 11.10.2022 adopted by the Italian Ministry of Research, CUP J97G22000390007.
P.S. acknowledges support from
ERC AdG NOQIA; MCIN/AEI (PGC2018-0910.13039/501100011033, CEX2019-000910-S/10.13039/50110 0011033, Plan National FIDEUA PID2019-106901GB-I00, Plan National STAMEENA PID2022-139099NB-I00 project funded by MCIN/AEI/10.13039/501100011033 and by the ``European Union NextGenerationEU/PRTR'' (PRTR-C17.I1), FPI); QUANTERA MAQS PCI2019-111828-2);  QUANTERA DYNAMITE PCI2022-132919 (QuantERA II Programme co-funded by European Union's Horizon 2020 program under Grant Agreement No 101017733), Ministry of Economic Affairs and Digital Transformation of the Spanish Government through the QUANTUM ENIA project call - Quantum Spain project, and by the European Union through the Recovery, Transformation, and Resilience Plan - NextGenerationEU within the framework of the Digital Spain 2026 Agenda; Fundaci\'{o} Cellex; Fundaci\'{o} Mir-Puig; Generalitat de Catalunya (European Social Fund FEDER and CERCA program, AGAUR Grant No. 2021 SGR 01452, QuantumCAT \ U16-011424, co-funded by ERDF Operational Program of Catalonia 2014-2020); Barcelona Supercomputing Center MareNostrum (FI-2023-2-0024); EU Quantum Flagship (PASQuanS2.1, 101113690); EU Horizon 2020 FET-OPEN OPTOlogic (Grant No 899794); EU Horizon Europe Program (Grant Agreement 101080086 - NeQST), ICFO Internal ``QuantumGaudi'' project; European Union's Horizon 2020 program under the Marie Sk{\l}odowska-Curie grant agreement No 847648;  ``La Caixa'' Junior Leaders fellowships, ``La Caixa'' Foundation (ID 100010434): CF/BQ/PR23/11980043. Views and opinions expressed are, however, those of the author(s) only and do not necessarily reflect those of the European Union, European Commission, European Climate, Infrastructure and Environment Executive Agency (CINEA), or any other granting authority.  Neither the European Union nor any granting authority can be held responsible for them. The work of AS was funded by the European Union - NextGenerationEU under the project NRRP “National Centre for HPC, Big Data and Quantum Computing (HPC)'' CN00000013 (CUP D43C22001240001) [MUR Decree n. 341- 15/03/2022] - Cascade Call launched by SPOKE 10 POLIMI: “CQEB” project.
C.V. is grateful to Federico Balducci for useful suggestions.

\end{acknowledgments}

\bibliography{references.bib}

\pagebreak
\widetext
\newpage

\section{Supplemental material - Renormalization Group Analysis of the Anderson Model on Random Regular Graphs}

\section*{$\beta$-function in two dimensions}
In the main text we have presented some general properties of the $\beta$-function. Here we conduct here a more detailed analysis of its behavior at the lower critical dimension $d=2$.

The $\beta$-function in $d=2$ is always negative  and it has a shallow fixed point at $D=1$ (see Eq.~(10) and Eq.~(13) in the main text)
\begin{equation}
\label{eq:beta_d2}
    \beta_2(D)=
    \begin{cases}
        \frac{1}{2}\ln D+O(1), & \text{if } D\ll 1\\
        -a (1-D)^2, & \text{if } D\simeq 1.
    \end{cases}
\end{equation}

From the numerics, we find $a\simeq 1$ (see Fig. 3 in the main text), which we will assume now to be the case, in agreement with sigma-model calculations, in particular Eq.~(22) in the main text.

\begin{figure}[t]
    \centering
    \includegraphics[width=0.6\linewidth]{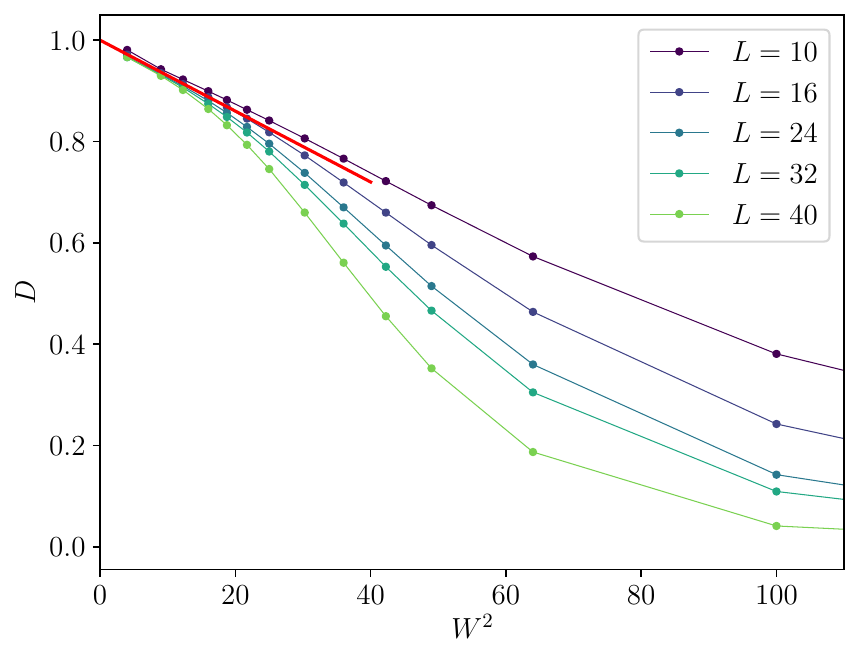}
    \caption{Fractal dimension for $d=2$ extracted from the participation entropy according to Eq.~(5) in the main text. It is clearly visible that near $D \simeq 1$ the dependence is of the form $D = 1 - a W^2$, as indicated by the red line.}
    \label{fig:d2_D_smallW}
\end{figure}

Let us consider the behavior at small $W$ (i.e. near $D=1$). Inserting $-(1-D)^{2}$ into r.h.s. of the RG Eq.~(7) in the main text we find: 
\begin{equation}
    d\left( \ln L^2 + \frac{1}{1-D} \right) = 0.
\end{equation}
This means that  
\begin{eqnarray}
\label{integral}
    \xi&=&L\exp\left(\frac{1}{2(1-D(W,L))}\right)\nonumber\\
    &=&\ell \exp\left(\frac{1}{2(1-D(W,\ell))} \right),
\end{eqnarray}
is constant along the RG trajectory which is fixed by initial conditions, {\it i.e.}\ by the value of r.h.s.\ of Eq.~(\ref{integral}) at the smallest length $L=\ell$ where the single-parameter scaling is still valid (an ultraviolet cutoff).
This is the localization length. To see its $W$ dependence at small $W$ we assume that:
\begin{equation}
    D(W,\ell)=1-(W/W_0)^2+O(W^3),
\end{equation}
as one can see in Fig.~\ref{fig:d2_D_smallW} (the constant $W_0$ depends on the cutoff $\ell$).

This could have been inferred from the fact that, at finite $L$, when $W\to 0$ all the observables are analytic in the variance and therefore must depend on $W^2$ analytically. This implies that, at small $W$, one obtains:
\begin{equation}
    \xi=\ell\exp\left(\frac{W_0^2}{2W^2}\right).
\end{equation}
This is in agreement with the well-known weak-localization result that in two dimensions $\ln(\xi/\ell)$ is proportional to the Drude conductivity and thus to the mean free path (MFP).  Indeed,  from a simple calculation of the decay rate of a wave packet with definite momentum (in the middle of the band), we have $\ell_{\mathrm{MFP}}=vt_{\mathrm{MFP}}$
\begin{equation}
    \frac{\hbar}{t_{\mathrm{MFP}}}=N\int d^2k'\delta(E_k-E_{k'})|\bra{k}\hat{V}\ket{k'}|^2\propto W^2,
\end{equation}
where $\hat{V}$ is the on-site potential and $\ket{k}$ is the plane wave of momentum $k$. This gives $\ell_{\mathrm{MFP}}\sim 1/W^2$ and therefore $\ln(\xi)\sim 1/W^2$, as seen for example in Ref.~\cite{lee1985disordered}.
\\

We now consider the behavior at large $W$. In this regime we have according to Eq.~(10) in the main text:
\begin{equation}
    \beta(D) \simeq \frac{1}{2}\ln D,
\end{equation}
which is compatible with a solution of the form  $D \sim (1/W)^L=\exp[-L/\xi]$, where 
\begin{equation}
    \xi \simeq \frac{1}{\ln W}.
\end{equation}

The complete dependence of $\xi$ on $W$, therefore, has to interpolate between $\xi \sim \exp(c/W^2)$, which is the \emph{weak localization regime} and $\xi \sim 1/\ln W$, which is the \emph{strong localization regime}. Therefore, the complete functional dependence should pass through a region of deceleration. We believe this has led to some claims in the literature that the scaling $\ln\xi\sim 1/W^\mu$ with $\mu$ close or even equal to 1~\cite{vidmar2023localization}.  
\\

Finally, we present the numerical results on the $L$-dependence of $D(L)$ and obtain the $\beta$-function for a two-dimensional system.  
The set of data on the $L$-dependence of $D(L)$ for different $W$ obtained from Eq.~(5) of the main text is presented in Fig.~2 of the main text, where the eigenfunction Shannon entropy is computed from Eq.~(10) in the main text using the eigenfunctions from the exact diagonalization of the Anderson model and averaging over disorder and eigenfunctions. From this set of data we obtain the plot $\beta(D)$ vs $D$ which is presented in Fig.~3 in the main text.

Remarkably, all the RG trajectories lie almost exactly on a single curve, just corroborating the single-parameter scaling as a very precise approximation in $d=2$.

\end{document}